\documentclass[sigconf, screen]{acmart}
\usepackage{graphicx}
\usepackage{subcaption}
\usepackage{tcolorbox}
\usepackage{bibunits}
\tcbuselibrary{breakable}

\AtBeginDocument{%
}

\setcopyright{acmlicensed}
\copyrightyear{2026}
\acmYear{2026}
\acmDOI{10.1145/3767308.3835017}

\acmConference[MM '26]{the 34th ACM International Conference on Multimedia}{November 10--14, 2026}{Rio de Janeiro, Brazil}

\acmISBN{}

\begin{document}
\begin{bibunit}[ACM-Reference-Format]

\title{ThinkOmni: A Reasoning-Driven Omni-Modal LLM Framework for Audio Forgery Detection and Localization}

\author{Yuxiong Xu}
\orcid{0000-0002-0514-3698}
\affiliation{%
	\institution{Guangdong Provincial Key Laboratory of Intelligent Information Processing, Shenzhen Key Laboratory of Media Security, \\ Shenzhen University}
	\city{Shenzhen}
	\country{China}}
\email{xuyuxiong2022@email.szu.edu.cn}

\author{Kaiqing Lin}
\orcid{0000-0002-5291-4635}
\affiliation{%
	\institution{Guangdong Provincial Key Laboratory of Intelligent Information Processing, Shenzhen Key Laboratory of Media Security, \\ Shenzhen University}
	\city{Shenzhen}
	\country{China}}
\email{linkaiqing2021@email.szu.edu.cn}

\author{Bin Li}
\authornote{Corresponding author.}
\orcid{0000-0002-2613-5451}
\affiliation{%
	\institution{Guangdong Provincial Key Laboratory of Intelligent Information Processing, Shenzhen Key Laboratory of Media Security, \\ Shenzhen University}
	\city{Shenzhen}
	\country{China}}
\email{libin@szu.edu.cn}

\author{Haodong Li}
\orcid{0000-0003-0532-9481}
\affiliation{%
	\institution{Guangdong Provincial Key Laboratory of Intelligent Information Processing, Shenzhen Key Laboratory of Media Security, \\ Shenzhen University}
	\city{Shenzhen}
	\country{China}}
\email{lihaodong@szu.edu.cn}

\author{Sheng Li}
\orcid{0009-0000-2761-1918}
\affiliation{%
	\institution{Afirstsoft Technology Group Co., Ltd.}
	\city{Shenzhen}
	\country{China}}
\email{admin@tenorshare.cn}

\renewcommand{\shortauthors}{Xu et al.}

\begin{abstract}
	Existing audio forgery detection and localization (AFDL) methods often overfit dataset-specific low-level artifacts, limiting their generalization to subtle, localized, and unseen manipulations. Recent audio large language model (ALLM)-based approaches cast AFDL as question answering but still model forensic evidence implicitly, without linking manipulation cues to predictions. To bridge this gap, we propose ThinkOmni, a reasoning-driven omni-modal large language model that jointly performs explicit forensic reasoning, spoofing detection, and temporal manipulation localization. To enable explicit reasoning supervision, we construct Forensic-Aware Chain-of-Thought (FACoT), a 100K-sample dataset with structured forensic evidence and reasoning annotations. Leveraging FACoT, we introduce Forensic-Aware Modality-Incremental Learning (FMIL), which progressively aligns semantic, acoustic, and spectral-visual representations with the LLM backbone to capture complementary forensic cues. We further propose Forensic-Consistent Multi-task Loss (FCML), which combines weighted cross-entropy with an adaptive localization loss to coordinate reasoning generation, spoofing detection, and temporal localization. Extensive experiments show that ThinkOmni achieves strong cross-dataset generalization in both detection and localization. Code, models, data, and inference examples are available at \url{https://beyond0814.github.io/ThinkOmni/}.
\end{abstract}



\maketitle

\begin{figure}[t]
\centering
\includegraphics[width=\linewidth]{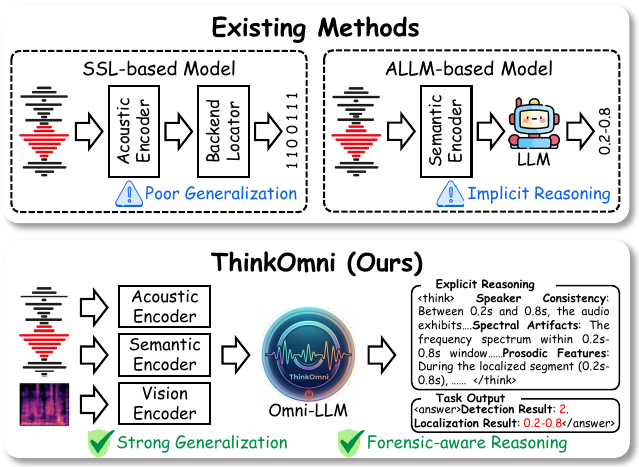}
\caption{Comparison of AFDL paradigms. SSL-based methods predict authenticity or temporal boundaries directly, whereas ALLM-based methods formulate AFDL as sequence generation but often lack explicit forensic reasoning supervision. ThinkOmni integrates semantic, acoustic, and spectral features for joint reasoning, detection, and localization.}
\label{fig:model_comparison}
\end{figure}

\section{Introduction}
Advances in generative models have expanded the scope of audio spoofing from fully synthetic speech to increasingly fine-grained partial manipulations \cite{Fun-Audio-Chat,Step-Audio2,AudioEditor,Li2024}. By altering only short temporal segments while preserving most of the original recording, partial deepfakes leave localized and less perceptible forensic traces, posing substantial challenges to forgery detection and temporal localization, especially under cross-dataset evaluation \cite{SurveyXu,HAD}.

Recent studies \cite{SpoofDiarization,TDAM,Chen2025} address AFDL through two paradigms: self-supervised learning (SSL)-based methods and audio large language model (ALLM)-based methods. 
Although they differ in formulation, both rely largely on \textit{implicit representations learned from training data rather than explicitly modeling how forensic cues support detection and localization}.
SSL-based methods \cite{BAM,CFPRF,Xu2024,Xu2025} fine-tune pre-trained acoustic encoders to capture manipulation artifacts and temporal boundaries, but may overfit to dataset-specific low-level artifacts (e.g., synthesis traces or manipulation patterns). 
ALLM-based methods \cite{DFALLM,ALLM4ADD} formulate AFDL as a question-answering task and leverage the prior knowledge of pre-trained ALLMs.
However, in the absence of explicit forensic reasoning, their predictions remain largely driven by implicit latent representations and learned correlations. Consequently, both paradigms exhibit limited generalization to unseen datasets \cite{partialspoof,Speech-Forensics,Llamapartialspoof}.

These limitations raise a key question: \textbf{can AFDL benefit from explicit forensic reasoning beyond implicit low-level features?} We argue that manipulation boundaries, speaker inconsistencies, and semantic--context mismatches can provide complementary and potentially more transferable evidence. Unlike conventional SSL-based acoustic models that mainly rely on feature matching, ALLMs possess strong reasoning and instruction-following capabilities. We therefore introduce Chain-of-Thought (CoT) \cite{wei2022chain} to decompose AFDL into intermediate reasoning steps and progressively analyze multi-modal forensic evidence \cite{FT-GRPO,Veritas,VLForgery}. This formulation encourages the model to connect observable cues with detection and localization targets instead of producing predictions solely from latent correlations.

In this paper, we propose ThinkOmni, a reasoning-driven omni-modal LLM built upon Qwen2.5-Omni \cite{Qwen2.5-Omni}. 
As illustrated in Figure~\ref{fig:model_comparison} (Bottom), ThinkOmni unifies explicit forensic reasoning, spoofing detection, and temporal manipulation localization within a single framework.
To support explicit forensic reasoning, we construct Forensic-Aware Chain-of-Thought (FACoT), a large-scale reasoning dataset for partially deepfake audio, comprising human-machine collaborative annotations.
Beyond forgery labels and temporal boundaries, FACoT provides structured supervision over semantic inconsistencies, acoustic artifacts, and temporal manipulation patterns, enabling the model to reason from diverse forensic evidence.
To effectively leverage FACoT and improve cross-dataset generalization, we further introduce a progressive training strategy and a task-specific loss function. 
Specifically, Forensic-Aware Modality-Incremental Learning (FMIL) progressively aligns semantic, acoustic, and spectral-visual representations with the LLM backbone to capture complementary and transferable forensic cues. 
Forensic-Consistent Multi-task Loss (FCML) combines weighted cross-entropy with an adaptive localization loss to jointly optimize spoofing detection and temporal localization.

Our main contributions are summarized as follows:
\begin{itemize}
	\item We propose \textbf{ThinkOmni}, a reasoning-driven omni-modal LLM that jointly performs explicit forensic reasoning, spoofing detection, and temporal manipulation localization within a unified AFDL framework.
	\item We construct \textbf{FACoT}, a large-scale 100K-sample dataset with structured reasoning annotations, establishing a new supervision paradigm for partially deepfake audio.
	\item We introduce a unified training strategy that couples \textbf{FMIL}-based progressive multi-modal alignment with \textbf{FCML}-guided detection–localization optimization to learn transferable forensic representations across datasets.
\end{itemize}

\begin{figure*}[ht]
	\centering
	\includegraphics[width=\textwidth]{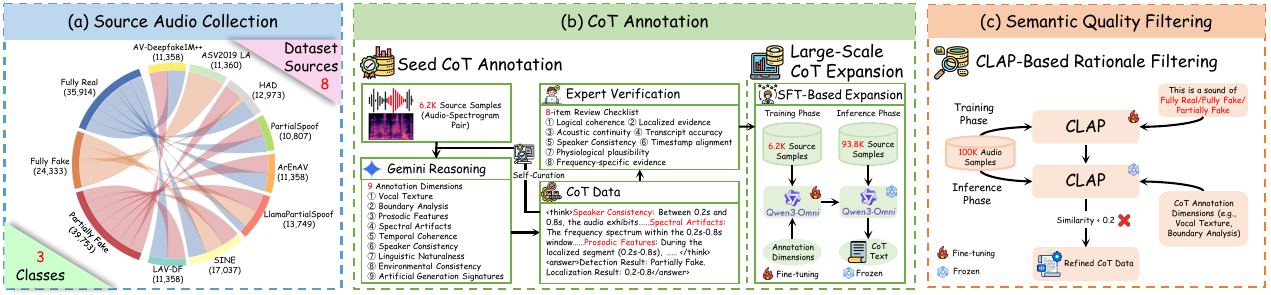}
	\caption{Construction pipeline of FACoT. (a) Audio selection from eight source datasets across three classes.  (b) Human-machine collaborative CoT annotation with SFT-based scaling. (c) CLAP-based filtering for audio-text consistency.}
	\label{fig:dataset_pipeline}
\end{figure*}

\section{Related Work}
\subsection{Audio Forgery Detection and Localization}
AFDL jointly assesses audio authenticity and localizes manipulated temporal segments.
Existing methods can be divided into self-supervised learning (SSL)-based and audio large language model (ALLM)-based approaches.

\noindent
\textbf{SSL-based Methods.}
SSL-based methods typically formulate temporal manipulation localization as either frame-level classification or boundary detection. Frame-level methods assign authenticity labels to short temporal units and derive manipulated regions from frame-wise predictions \cite{AGO,GNCL}. Representative approaches include MRM \cite{partialspoof}, which combines SSL representations with multi-resolution modeling to capture manipulations at different temporal scales; TDL \cite{TDL}, which exploits embedding similarity for frame-level discrimination; and PET \cite{PET}, which models high-frequency components and temporal consistency to expose splicing artifacts.

Boundary-based methods identify transitions between genuine and manipulated regions. CFPRF \cite{CFPRF} progressively refines coarse temporal proposals to obtain precise manipulation boundaries, whereas BAM \cite{BAM} employs boundary-aware attention to improve localization accuracy. Despite their effectiveness, these methods rely predominantly on low-level acoustic artifacts and are therefore susceptible to dataset- and generator-specific patterns, resulting in limited cross-dataset generalization. Moreover, their predictions are produced through implicit feature matching, without explicit reasoning over the forensic evidence underlying the decisions.

\noindent
\textbf{ALLM-based Methods.}
ALLMs jointly encode audio and textual instructions, enabling instruction-following across diverse audio understanding tasks \cite{Qwen-audio,Qwen2-audio,Qwen2.5-Omni}. However, their application to audio forensics remains underexplored. Recent studies cast AFDL as a question-answering task within the ALLM framework. DFALLM \cite{DFALLM} improves generalization through multi-task adaptation of the audio encoder and language model. HoliAntiSpoof \cite{HoliAntiSpoof} jointly models attack identification, temporal localization, and semantic impact assessment. PELM \cite{PELM} further incorporates frame-level probabilities from conventional detectors as auxiliary evidence for forgery detection and localization.

Although these methods extend AFDL beyond conventional classification, their decisions remain largely driven by latent correlations. They neither organize forensic evidence nor supervise the reasoning process linking manipulation cues to detection and localization, which can limit generalization to unseen datasets.

\subsection{Chain-of-Thought}
Chain-of-Thought (CoT) models reasoning steps and improves performance on complex reasoning tasks \cite{wei2022chain,zhouleast}. This capability is well suited to multimedia forensics, where reliable decisions require both manipulation detection and evidence-grounded analysis. Studies have introduced CoT supervision into visual forensics \cite{lin2025seeing,Veritas}. EDVD\_LLaMA \cite{EDVD_LLaMA} incorporates facial cues into multi-modal CoT for spatio-temporal localization, while HEIE \cite{HEIE} decomposes forged-image detection into progressively harder subtasks.

In audio forensics, FT-GRPO \cite{FT-GRPO} introduces frequency--time CoT rationales for spoofing analysis. However, its reasoning is limited to time--frequency artifacts and overlooks the generalization properties of acoustic and semantic encoders \cite{DFALLM}. As a result, semantic inconsistencies, acoustic artifacts, and temporal manipulation patterns are not modeled. To bridge this gap, we develop a structured CoT annotation pipeline for AFDL that integrates semantic, acoustic, and temporal evidence, providing supervision for forensic reasoning, spoofing detection, and temporal localization.

\section{Preliminary}
\label{sec:preliminary}

\subsection{Task Definition}
\label{subsec:task_def}

Given an audio waveform $\boldsymbol{A}$, its spectrogram $\boldsymbol{S}$, and a forensic instruction $\boldsymbol{I}$, we define
$\boldsymbol{X}=(\boldsymbol{A},\boldsymbol{S},\boldsymbol{I})$.
ThinkOmni generates a structured output
$\boldsymbol{Y}=(\boldsymbol{r},c,\boldsymbol{z})$, where
$\boldsymbol{r}$ is the forensic evidence reasoning sequence,
$c\in\{0,1,2\}$ denotes fully real, fully fake, and partially fake audio, respectively, and
$\boldsymbol{z}$ is the timestamp-token sequence.
The structured output is serialized into
$(y_1,\ldots,y_N)$ and generated autoregressively as:
\begin{equation}
	P_{\theta}(\boldsymbol{Y}\mid\boldsymbol{X})
	=
	\prod_{n=1}^{N}
	P_{\theta}(y_n\mid y_{<n},\boldsymbol{X}),
\end{equation}
where $P_{\theta}$ is the conditional distribution parameterized by $\theta$, $N$ is the output length, and $y_{<n}=(y_1,\ldots,y_{n-1})$.
At inference, the predicted timestamp tokens are parsed into temporal intervals:
\begin{equation}
	\hat{\mathcal{B}}
	=
	g(\hat{\boldsymbol{z}})
	=
	\{[\hat{s}_j,\hat{e}_j]\}_{j=1}^{\hat{K}},
\end{equation}
where $g(\cdot)$ is the token-to-interval parser,
$\hat{s}_j$ and $\hat{e}_j$ are the predicted start and end times, and
$\hat{K}$ is the number of predicted segments.
The ground-truth intervals are
$\mathcal{B}=\{[s_k,e_k]\}_{k=1}^{K}$,
where $s_k$, $e_k$, and $K$ denote the corresponding ground-truth quantities.

\subsection{Forensic-Aware Chain-of-Thought Dataset}
Partially deepfake audio modifies only selected temporal segments while preserving most of the original recording, producing subtler and more localized forensic traces than fully synthetic audio \cite{psd_survey_yi}. Existing datasets, however, are primarily designed for direct supervision and typically provide only forgery labels and temporal boundaries, without structured annotations explaining the underlying forensic evidence. To address this limitation, we develop a cost-effective human-machine collaborative pipeline to construct Forensic-Aware Chain-of-Thought (FACoT), a large-scale dataset for partially deepfake audio with structured annotations of semantic inconsistencies, acoustic artifacts, and temporal manipulation patterns, as illustrated in Figure~\ref{fig:dataset_pipeline}.

FACoT is constructed in three stages:
(1) \textbf{Source Audio Collection}, which aggregates 100K samples from eight public datasets;
(2) \textbf{CoT Annotation}, which combines expert-guided seed annotation with model-based large-scale expansion; and
(3) \textbf{Semantic Quality Filtering}, which removes audio-inconsistent reasoning dimensions using contrastive language--audio pretraining (CLAP).

\noindent
\textbf{Source Audio Collection.}
Real-world audio forgeries span diverse generation methods, editing operations, speakers, and acoustic conditions, whereas individual datasets cover only a limited subset of these variations. We therefore aggregate 100K samples from eight representative public datasets, as shown in Figure~\ref{fig:dataset_pipeline}(a). Specifically, the collection includes ASVspoof 2019 LA (11,360) \cite{ASV2019}, HAD (12,973) \cite{HAD}, PartialSpoof (10,807) \cite{partialspoof}, LAV-DF (11,358) \cite{LAV-DF}, ArEnAV (11,358) \cite{ArEnAV}, LlamaPartialSpoof (13,749) \cite{Llamapartialspoof}, SINE (17,037) \cite{SINE}, and AV-Deepfake1M++ (11,358) \cite{Av-deepfake1m++}. The resulting collection comprises 35,914 fully real, 24,333 fully fake, and 39,753 partially fake samples, covering diverse spoofing mechanisms and acoustic conditions. This broad coverage provides a representative foundation for constructing forensic reasoning annotations.

\begin{figure*}[htbp]
	\centering
	\includegraphics[width=\textwidth]{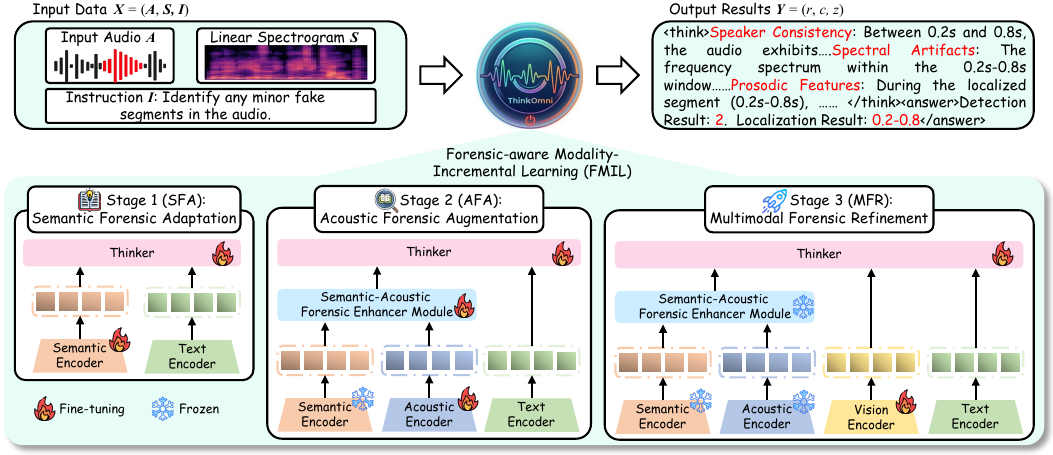}
	\caption{Overview of ThinkOmni. We propose a progressive forensic-aware modality-incremental learning (FMIL) strategy that incrementally aligns the semantic, acoustic, and vision encoders with the Thinker backbone, ensuring that multi-modal forensic features generalize effectively without disrupting previously learned alignments.}
	\label{fig:framework}
\end{figure*}

\noindent
\textbf{CoT Annotation.}
To balance annotation quality and scalability, we adopt a two-step human-machine collaborative procedure, as shown in Figure~\ref{fig:dataset_pipeline}(b). We first construct a 6.2K-sample seed set through stratified sampling across datasets and classes. The resulting annotations are then used to adapt Qwen3-Omni \cite{Qwen3-omni}, which generates CoT annotations for the remaining 93.8K samples.

\textbf{1) Seed CoT Annotation} 
We select 6.2K audio samples from the eight source datasets to construct the seed set. Each sample is provided to Gemini-3-Pro \cite{gemini}, together with its spectrogram, forgery label, and temporal boundaries, to generate an initial reasoning trace. The annotation schema contains nine forensic dimensions organized into three hierarchical levels: low-level acoustic anomalies, including vocal texture, spectral artifacts, and generation signatures; mid-level temporal discontinuities, including boundary characteristics and temporal coherence; and high-level contextual inconsistencies, including prosody, speaker consistency, linguistic naturalness, and environmental consistency.

The annotations are refined through two quality-control procedures. First, \textit{Self-Curation} verifies each rationale against the forgery label and temporal boundaries. Second, during \textit{Expert Verification}, a forensic expert assesses each annotation using an eight-item checklist covering semantic and logical correctness, cross-modal temporal alignment, and acoustic and physical grounding. The checklist evaluates logical coherence, transcript accuracy, localized evidence, timestamp alignment, acoustic continuity, speaker consistency, physiological plausibility, and frequency-level justification.

\textbf{2) Large-Scale CoT Expansion.}
To scale annotation, we fine-tune Qwen3-Omni on the 6.2K seed samples and spectrograms using low-rank adaptation \cite{Lora}. The adapted model then generates structured reasoning annotations for the remaining 93.8K samples, yielding 100K CoT-annotated samples.

\noindent
\textbf{Semantic Quality Filtering.}
Automatically generated rationales may contain content weakly grounded in the audio, potentially introducing noisy supervision during training. We therefore apply a CLAP-based semantic consistency filter to reasoning dimensions, as shown in Figure~\ref{fig:dataset_pipeline}(c) \cite{CLAP}. To adapt CLAP to audio-forensic semantics, we fine-tune it using class-aware audio-text pairs constructed from the 100K samples, with prompts corresponding to fully real, fully fake, and partially fake audio.

For each sample, we compute the similarity between its audio embedding and text embedding of each reasoning dimension. Dimensions with similarity scores below 0.2 are removed, while the remaining dimensions are retained as supervision. This dimension-level filtering preserves all 100K audio samples while discarding weakly grounded rationale components. The resulting FACoT dataset provides structured CoT annotations with improved audio--text consistency for training reasoning-driven audio forensic models.

\section{Method}
\subsection{Overview}
We propose ThinkOmni, a reasoning-driven omni-modal framework built on Qwen2.5-Omni \cite{Qwen2.5-Omni} for audio forensics. As shown in Figure~\ref{fig:framework}, ThinkOmni retains the semantic encoder, vision encoder, and Thinker backbone, while incorporating an acoustic encoder and a Semantic-Acoustic Forensic Enhancer (SAFE) to capture complementary low-level forensic cues.

To facilitate stable multi-modal adaptation, we introduce Forensic-Aware Modality-Incremental Learning (FMIL), a progressive training strategy comprising Semantic Forensic Adaptation (SFA), Acoustic Forensic Augmentation (AFA), and Multi-modal Forensic Refinement (MFR). These stages progressively integrate semantic, acoustic, and spectral-visual representations while reducing interference among heterogeneous modalities. We further propose Forensic-Consistent Multi-task Loss (FCML), which combines weighted cross-entropy with an adaptive localization loss to jointly optimize spoofing detection and temporal localization.

\begin{figure}[t]
	\centering
	\includegraphics[width=\linewidth]{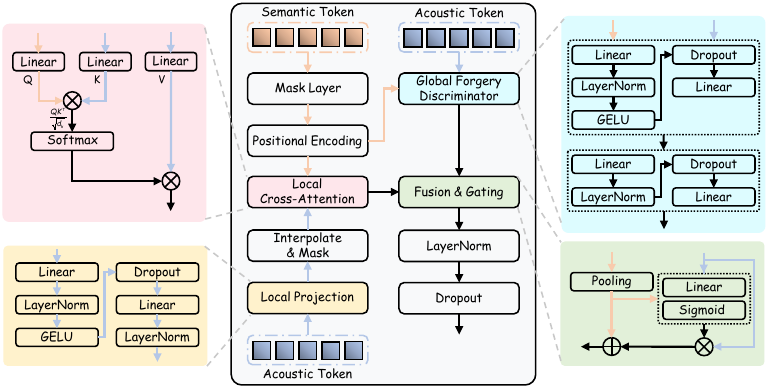}
	\caption{Architecture of SAFE. The cross-attention module captures local forensic cues, while the forgery discriminator models global forensic context.}
	\label{fig:fusion_module}
\end{figure}

\subsection{Forensic-aware Modality-Incremental Learning}
\noindent
\textbf{Motivation.}
Cross-dataset generalization in audio forensics is hindered by dataset-specific artifacts and heterogeneous forensic cues across modalities. To address this issue, FMIL progressively incorporates semantic, acoustic, and spectral-visual evidence through three stages. Semantic Forensic Adaptation (SFA) first establishes transferable semantic reasoning from speech content and speaker information. Acoustic Forensic Augmentation (AFA) then introduces fine-grained acoustic evidence to capture subtle manipulation artifacts. Finally, Multi-modal Forensic Refinement (MFR) integrates spectrogram-based visual cues for cross-modal verification. This progressive semantic-to-multi-modal training reduces modality interference and reliance on dataset-specific shortcuts.

\noindent
\textbf{Semantic Forensic Adaptation.}
SFA establishes the semantic reasoning foundation of ThinkOmni. Given an audio sample $\boldsymbol{A}$ and textual instruction $\boldsymbol{I}$, the semantic encoder extracts speech representations, while the text encoder encodes the instruction. These features are aligned with the Thinker backbone through supervised fine-tuning. By prioritizing semantic reasoning before introducing low-level artifacts, SFA captures contextual and speaker-related inconsistencies that are transferable across manipulation methods.

\noindent
\textbf{Acoustic Forensic Augmentation.}
Semantic representations capture high-level inconsistencies but may overlook subtle artifacts, such as phase discontinuities and temporal jitter. AFA therefore introduces a dedicated acoustic encoder while freezing the semantic encoder to preserve the learned semantic representations.

To align semantic and acoustic evidence, we propose the SAFE module (Figure~\ref{fig:fusion_module}), which comprises a local cross-attention branch for fine-grained cue interaction and a global forgery discriminator for long-range forensic modeling. During AFA, only the acoustic encoder, SAFE, and Thinker backbone are updated, allowing acoustic cues to complement semantic reasoning without disrupting the established representations.

\noindent
\textbf{Multi-modal Forensic Refinement.}
Certain forgery artifacts are more distinguishable in spectrograms than in raw waveforms. MFR therefore introduces a vision encoder to capture spectral-visual evidence while freezing the semantic encoder, acoustic encoder, and SAFE. Only the vision encoder and Thinker backbone are optimized, aligning visual cues with the established forensic reasoning space. By integrating semantic, acoustic, and spectral-visual evidence, MFR enables more reliable cross-modal verification and improves cross-dataset generalization.

\subsection{Forensic-Consistent Multi-task Loss}
Reasoning tokens dominate the output sequence, biasing standard cross-entropy optimization toward the reasoning task. To mitigate this imbalance, FCML combines weighted cross-entropy with an adaptive localization loss to coordinate reasoning, detection, and localization.

\noindent
\textbf{Weighted Cross-Entropy Loss.}
We apply weighted cross-entropy $\mathcal{L}_{wce}$ to the structured output sequence $\boldsymbol{Y}=(y_1,\dots,y_N)$ conditioned on the multi-modal input $\boldsymbol{X}$:
\begin{equation}
	\mathcal{L}_{wce} = - \sum_{n=1}^{N} \omega(y_n) \log P_{\theta}(y_n \mid y_{<n}, \boldsymbol{X}),
\end{equation}
where $N$ denotes the sequence length and $\theta$ represents the model parameters.
The token weight $\omega(y_n)$ dynamically adjusts based on its structural role:
\begin{itemize}
	\item \textbf{Reasoning Token:} $\omega_{think} = 0.2$, reducing the influence of reasoning-token gradients and preventing them from dominating the optimization process during forensic learning.
	\item \textbf{Detection Token:} We use a class-prior-aware weight $\omega_{det}=\alpha_{det}\cdot\omega_{cls}$, where $\alpha_{det}=0.2$ and $\omega_{cls}=(0.36,0.24,0.40)$ correspond to fully real, fully fake, and partially fake samples, respectively, following the class proportions in FACoT.
	\item \textbf{Localization Token:} $\omega_{loc} = 0.6$, enforcing strict adherence to precise temporal boundaries.
\end{itemize}

\begin{table*}[t]
	\centering
	\caption{Comparison of ThinkOmni with SOTA methods for intra- and cross-dataset spoofing detection.}
	\label{table:detection}
	\resizebox{\textwidth}{!}{
		\begin{tabular}{l cc cc cc cc cc cc cc cc cc cc cc}
			\toprule
			&
			& \multicolumn{15}{c}{\textbf{Intra-Dataset}} 
			& \multicolumn{6}{c}{\textbf{Cross-Dataset}} \\
			\cmidrule(lr){2-17} \cmidrule(lr){18-23}
			\raisebox{-1ex}[0pt][0pt]{\textbf{Method}} 
			& \multicolumn{2}{c}{PS} 
			& \multicolumn{2}{c}{HAD} 
			& \multicolumn{2}{c}{LAV-DF} 
			& \multicolumn{2}{c}{SINE} 
			& \multicolumn{2}{c}{LPS} 
			& \multicolumn{2}{c}{ArEnAV} 
			& \multicolumn{2}{c}{AV-1M++} 
			& \multicolumn{2}{c}{Avg.} 
			& \multicolumn{2}{c}{ADD} 
			& \multicolumn{2}{c}{SF} 
			& \multicolumn{2}{c}{Avg.} \\
			\cmidrule(lr){2-3} \cmidrule(lr){4-5} \cmidrule(lr){6-7}
			\cmidrule(lr){8-9} \cmidrule(lr){10-11} \cmidrule(lr){12-13}
			\cmidrule(lr){14-15} \cmidrule(lr){16-17}
			\cmidrule(lr){18-19} \cmidrule(lr){20-21} \cmidrule(lr){22-23} 
			& ACC & F1 & ACC & F1 & ACC & F1 & ACC & F1 
			& ACC & F1 & ACC & F1 & ACC & F1 
			& ACC & F1 
			& ACC & F1 & ACC & F1 & ACC & F1 \\
			\midrule
			
			\multicolumn{23}{l}{\textit{\textbf{SSL-based Method}}} \\
			W2V2-AASIST \cite{tak2022automatic}
			& 89.80 & 89.13 & 99.21 & 99.40 & 90.79 & 90.84 & 78.39 & 78.27 
			& 85.74 & 85.99 & 96.42 & 96.43 & 89.63 & 89.76 
			& 90.00 & 89.97 
			& 60.35 & 61.25 & 21.90 & 28.81 & 41.13 & 45.03 \\
			
			W2V2-Conformer \cite{rosello2023conformer}
			& 91.79 & 91.70 & 99.32 & 99.47 & 95.58 & 95.59 & 82.60 & 82.54 
			& 88.38 & 88.81 & 96.89 & 96.89 & \underline{93.01} & \underline{93.07}
			& 92.51 & 92.58 
			& 66.72 & 65.90 & 26.72 & 34.92 & 46.72 & 50.41 \\
			
			TCM \cite{truong2024temporal}
			& 93.08 & 92.80 & \underline{99.44} & \underline{99.56} & 92.28 & 92.17 & \underline{86.72} & \underline{86.79} 
			& \underline{90.47} & \underline{90.48} & \underline{97.22} & \underline{97.23} & 90.68 & 90.40 
			& \underline{92.84} & \underline{92.78}
			& 67.41 & 64.98 & 23.51 & 33.96 & 45.46 & 49.47 \\
			
			XLSR-SLS \cite{zhang2024audio}
			& 90.16 & 89.80 & \textbf{99.58} & \textbf{99.67} & 95.01 & 95.01 & 82.88 & 82.86 
			& 87.74 & 88.21 & 96.94 & 96.95 & 92.26 & 92.32 
			& 92.08 & 92.12 
			& 68.96 & 67.59 & 20.57 & 28.04 & 44.77 & 47.82 \\
			
			Nes2Net-X \cite{liu2025nes2net}
			& 86.77 & 85.76 & 99.06 & 99.42 & 96.16 & 96.16 & \textbf{87.86} & \textbf{87.90} 
			& 88.39 & 87.97 & \textbf{97.52} & \textbf{97.52} & 92.87 & 92.88 
			& 92.66 & 92.52 
			& 71.02 & 67.61 & 13.68 & 21.18 & 42.35 & 44.40 \\
			
			\midrule
			\multicolumn{23}{l}{\textit{\textbf{ALLM-based Method}}} \\
			ALLM4ADD \cite{ALLM4ADD} 
			& \textbf{96.48} & \textbf{96.48} & 98.39 & 99.15 & 95.89 & 95.90 & 62.73 & 60.45 
			& 90.07 & 90.05 & 94.88 & 94.88 & 90.04 & 90.05 
			& 89.78 & 89.57 
			& 72.61 & 73.79 & 51.96 & 55.22 & 62.29 & 64.51 \\
			
			Qwen2-Audio \cite{Qwen2-audio} 
			& 84.21 & 83.77 & 90.04 & 94.60 & \underline{96.92} & \underline{96.92} & 59.97 & 49.08 
			& 65.75 & 62.15 & 92.14 & 92.14 & 91.36 & 91.37 
			& 82.91 & 81.43 
			& 69.18 & 69.96 & \textbf{83.15} & \underline{85.56} & \underline{76.17} & \underline{77.76} \\
			
			Qwen2.5-Omni-3B \cite{Qwen2.5-Omni} 
			& 87.31 & 87.07 & 90.13 & 94.68 & 96.73 & 96.74 & 72.99 & 73.42 
			& 82.51 & 82.48 & 92.08 & 92.08 & 89.93 & 89.98 
			& 87.38 & 88.06 
			& \underline{75.33} & \underline{78.56} & 47.05 & 51.17 & 61.19 & 64.87 \\
			
			Qwen2.5-Omni-7B \cite{Qwen2.5-Omni} 
			& 81.15 & 80.68 & 93.58 & 96.54 & 93.32 & 93.33 & 63.15 & 59.35 
			& 64.78 & 63.10 & 90.44 & 90.45 & 84.68 & 85.05 
			& 81.59 & 81.21 
			& 75.17 & 78.32 & 62.05 & 72.14 & 68.61 & 75.23 \\
			
			ThinkOmni (Ours)
			& \underline{93.87} & \underline{93.87} & 98.23 & 98.99 & \textbf{99.46} & \textbf{99.46} & 81.96 & 81.41 
			& \textbf{90.64} & \textbf{90.59} & 96.51 & 96.51 & \textbf{95.24} & \textbf{95.24} 
			& \textbf{93.70} & \textbf{93.72} 
			& \textbf{78.87} & \textbf{80.94} & \underline{82.61} & \textbf{89.35} & \textbf{80.74} & \textbf{85.15} \\
			
			\bottomrule
		\end{tabular}}
\end{table*}

\noindent
\textbf{Adaptive Localization Loss.}
To accurately localize multiple manipulated segments within an audio sample, let $\mathcal{B} = \{\boldsymbol{b}_k\}_{k=1}^K$ denote the set of $K$ ground-truth temporal intervals, where $\boldsymbol{b}_k=[s_k,e_k]$. The model predicts $\hat{\mathcal{B}} = \{\hat{\boldsymbol{b}}_j\}_{j=1}^{\hat{K}}$, where $\hat{\boldsymbol{b}}_j=[\hat{s}_j,\hat{e}_j]$ and $\hat{K}$ denotes the number of predicted intervals. The predicted and ground-truth intervals are matched according to their token assignments. Based on the utterance-level label $c\in\{0,1,2\}$, denoting fully real, fully fake, and partially fake samples, respectively, the condition-adaptive localization loss is defined as follows:
\begin{equation}
	\mathcal{L}_{loc} = 
	\begin{cases}
		\lambda_{fr}\,\frac{1}{\max(1, \hat K)}\sum_{j=1}^{\hat{K}}\text{Smooth}_{L1}(\hat{\boldsymbol{b}}_j) & c=0, \\
		0 & c=1, \\
		\lambda_{pf}\,\frac{1}{\max(1, \hat K)}\sum_{j=1}^{\hat{K}}\mathcal{L}_{reg}(\hat{\boldsymbol{b}}_j, \boldsymbol{b}_j) & c=2,
	\end{cases}
\end{equation}
where $\lambda_{fr}=0.3$ and $\lambda_{pf}=0.5$ are empirically set. For fully real samples ($c=0$), all $\hat{K}$ predicted boundaries are constrained to zero. For fully fake samples ($c=1$), boundary regression is omitted because the entire utterance is manipulated.

For partially fake samples ($c=2$), we adopt a hybrid regression loss $\mathcal{L}_{reg}$ that jointly enforces temporal overlap and boundary coordinate accuracy for each segment pair:
\begin{equation}
	\mathcal{L}_{reg}(\hat{\boldsymbol{b}}_j, \boldsymbol{b}_j) = 1 - \text{IoU}(\hat{\boldsymbol{b}}_j, \boldsymbol{b}_j) + \text{Smooth}_{L1}(\hat{\boldsymbol{b}}_j - \boldsymbol{b}_j).
\end{equation}

Here, $\text{IoU}(\cdot, \cdot)$ measures the 1D temporal Intersection over Union for the $k$-th segment, defined as:
\begin{equation}
	\text{IoU}(\hat{\boldsymbol{b}}_j, \boldsymbol{b}_j) = \frac{H_j}{(\hat{e}_j-\hat{s}_j) + (e_j-s_j) - H_j + \epsilon},
\end{equation}
where $H_j = \max(0, \min(\hat{e}_j, e_j) - \max(\hat{s}_j, s_j))$ denotes the temporal intersection, and $\epsilon = 10^{-8}$ ensures numerical stability. 
Additionally, $\operatorname{Smooth}_{L1}$ penalizes boundary-coordinate errors element-wise over $d \in \{\hat{s}_j-s_j, \hat{e}_j-e_j\}$. For fully real samples, the same loss is applied between each predicted interval $\hat{\boldsymbol{b}}_j$ and $\mathbf{0}$:
\begin{equation}
	\text{Smooth}_{L1}(d) = 
	\begin{cases}
		0.5 d^2 & |d| < 1, \\
		|d| - 0.5 & \text{otherwise}.
	\end{cases}
\end{equation}

\noindent
\textbf{Overall Loss.}
The overall training objective is
\begin{equation}
	\mathcal{L}_{total}
	=
	\mathcal{L}_{wce}
	+
	\lambda_{loc}\mathcal{L}_{loc},
\end{equation}
where $\lambda_{loc}=0.5$ balances structured sequence generation and
temporal boundary supervision.

\section{Experiments}
\subsection{Experimental Setup}
\noindent
\textbf{Datasets.}
ThinkOmni and all baselines are trained on the same 100K-sample FACoT pool, which combines eight public datasets: ASVspoof 2019 LA (19LA) \cite{ASV2019}, HAD \cite{HAD}, PartialSpoof (PS) \cite{partialspoof}, LAV-DF \cite{LAV-DF}, ArEnAV \cite{ArEnAV}, LlamaPartialSpoof (LPS) \cite{Llamapartialspoof}, SINE \cite{SINE}, and AV-Deepfake1M++ (AV-1M++) \cite{Av-deepfake1m++}. Baselines use only the labels or temporal boundaries required by their original objectives, while structured reasoning annotations are reserved for ThinkOmni and its reasoning-based variants.

For intra-dataset evaluation, we use non-overlapping test samples from the eight source datasets. ADD 2023 Track 2 (ADD) \cite{yi2023add} and Speech-Forensics (SF) \cite{Speech-Forensics} are used for cross-dataset evaluation. Since PS is derived from 19LA, fully fake 19LA test samples are assigned to PS to avoid duplication. As AV-1M++ lacks test labels, its development set is used for evaluation.

\noindent
\textbf{Comparison Methods.}
Using the common FACoT training protocol described above, we compare ThinkOmni with publicly reproducible SSL-based detection and localization methods, including W2V2-AASIST \cite{tak2022automatic}, W2V2-Conformer \cite{rosello2023conformer}, TCM \cite{truong2024temporal}, XLSR-SLS \cite{zhang2024audio}, Nes2Net-X \cite{liu2025nes2net}, MRM \cite{partialspoof}, TDL \cite{TDL}, BAM \cite{BAM}, and CFPRF \cite{CFPRF}. We also include the ALLM-based AFDL method ALLM4ADD \cite{ALLM4ADD} and representative general-purpose audio LLMs, including Qwen-Audio \cite{Qwen-audio}, Qwen2-Audio \cite{Qwen2-audio}, and Qwen2.5-Omni-3B/7B \cite{Qwen2.5-Omni}. For the acoustic-encoder ablation, we evaluate Wav2Vec2-XLSR-300M (XLSR-300M) \cite{XLSR}, Wav2Vec2-XLSR-1B (XLSR-1B) \cite{XLSR}, and Wav2Vec2-BERT (BERT) \cite{wav2vec2}. Reasoning quality is assessed using Qwen3.5-Omni (Qwen) \cite{Qwen3.5-Omni}, GPT-Audio (GPT)\footnote{https://developers.openai.com/api/docs/models/gpt-audio}, and MiMo-V2.5 (MiMo)\footnote{https://mimo.xiaomi.com/mimo-v2-5} as MLLM judges, together with human evaluation.

\noindent
\textbf{Evaluation Metrics.}
For spoof detection, we report accuracy (ACC) and F1-score (F1) to assess overall performance.
For temporal manipulation localization, we adopt mean Average Precision (mAP) over temporal IoU thresholds $[0.5:0.05:0.95]$ \cite{CFPRF}.
For reasoning evaluation, we use ROUGE\_L \cite{Rouge}, BLEU-4 \cite{Bleu}, METEOR \cite{Meteor}, and cosine semantic similarity (CSS).
These metrics measure similarity between generated and reference texts from complementary perspectives, including longest common subsequence, n-gram overlap, synonym matching, and semantic similarity.
Best and second-best results are highlighted in \textbf{bold} and \underline{underlined}.

\noindent
\textbf{Implementation Details.}
We implement ThinkOmni in PyTorch using ms-swift \cite{zhao2025swift}. Compatible ALLM models are fine-tuned with LoRA on linear layers, using $r=8$, $\alpha=32$, and dropout $0.05$. Each FMIL stage is trained for one epoch with learning rates of $1\times10^{-4}$ for the Thinker backbone and $1\times10^{-5}$ for the ViT and aligner. Additional settings are provided in the supplementary material.

\begin{table*}[t]
	\centering
	\caption{Comparison of ThinkOmni with SOTA methods for intra- and cross-dataset temporal localization.}
	\label{table:localization}
	\footnotesize
	\resizebox{\textwidth}{!}{
	\begin{tabular}{l cccccccc ccc}
		\toprule
		\raisebox{-2.5ex}[0pt][0pt]{\textbf{Method}} & \multicolumn{8}{c}{\textbf{Intra-Dataset}} & \multicolumn{3}{c}{\textbf{Cross-Dataset}} \\
		\cmidrule(lr){2-9} \cmidrule(lr){10-12}
		& PS & HAD & LAV-DF & SINE & LPS & ArEnAV & AV-1M++ & \textbf{Avg.} & ADD & SF & \textbf{Avg.} \\
		\midrule
		\multicolumn{12}{l}{\textit{\textbf{SSL-based Method}}} \\
		MRM \cite{partialspoof}  & 31.93 & 88.12 & 81.73 & 24.07 & 32.80 & \underline{92.09} & 79.04 & 61.40 & 0.62 & 0.05 & 0.34 \\
		TDL \cite{TDL}& \textbf{80.92} & 79.67 & 87.67 & 58.16 & 78.63 & 86.79 & 78.31 & 78.59 & 59.14 & 2.74 & 30.94 \\
		BAM \cite{BAM} & 51.24 & \textbf{98.65} & 88.72 & 62.22 & \underline{79.11} & 86.03 & \textbf{93.61} & 79.94 & 0.12 & 4.40 & 2.26 \\
		CFPRF \cite{CFPRF}  & 56.45 & 93.18 & 81.92 & 66.09 & 66.35 & 85.33 & 64.81 & 73.45 & 1.27 & 3.39 & 2.33 \\
		\midrule
		\multicolumn{12}{l}{\textit{\textbf{ALLM-based Method}}} \\
		Qwen-Audio \cite{Qwen-audio} & 67.08 & 33.22 & 63.82 & 41.09 & 60.23 & 61.84 & 48.02 & 53.61 & 32.41 & 1.59 & 17.00 \\
		Qwen2-Audio \cite{Qwen2-audio}& 67.80 & 85.10 & 88.54 & 69.04 & 64.05 & 75.32 & 68.43 & 74.04 & 67.92 & \underline{50.77} & \underline{59.35} \\
		Qwen2.5-Omni-3B \cite{Qwen2.5-Omni}& 77.42 & 78.83 & 90.85 & 80.45 & 73.48 & 87.72 & 76.40 & 80.74 & \underline{73.70} & 12.08 & 42.89 \\
		Qwen2.5-Omni-7B \cite{Qwen2.5-Omni}& \underline{77.77} & 84.86 & \underline{93.30} & \textbf{94.18} & 67.27 & 90.27 & 78.19 & \underline{83.69} & 68.77 & 43.04 & 55.91 \\
		ThinkOmni (Ours) & 74.20 & \underline{94.05} & \textbf{96.19} & \underline{84.34} & \textbf{79.14} & \textbf{96.69} & \underline{91.72} & \textbf{88.05} & \textbf{74.36} & \textbf{74.98} & \textbf{74.67} \\
		\bottomrule
	\end{tabular}}
\vspace{-2mm}
\end{table*}

\subsection{Detection Evaluation}
We compare ThinkOmni with state-of-the-art (SOTA) SSL- and ALLM-based methods under intra- and cross-dataset settings.
All methods are retrained under the same setup for fairness.

\noindent
\textbf{Intra-dataset Performance.}
As shown in Table~\ref{table:detection}, SSL-based methods achieve average ACC and F1 scores of approximately 90\%--93\%, with only modest variation across architectures. Despite their different designs, most methods rely on similar large-scale acoustic encoders, such as Wav2Vec 2.0 \cite{XLS-R} and WavLM-Large \cite{Wavlm}, each containing roughly 300M parameters. Their comparable performance highlights the strength of domain-specific acoustic representations for intra-dataset detection, with remaining differences arising from downstream architectures and optimization objectives. Comparisons with ALLM-based methods should also consider differences in model scale and training paradigms.

ThinkOmni further outperforms all baselines, achieving 93.70\% ACC and 93.72\% F1. 
Although the gains over the strongest SSL baselines are modest, the result highlights the benefit of jointly modeling multi-dimensional cues and maintains a clear advantage over the ALLM-based baselines.

\noindent
\textbf{Cross-dataset Performance.}
As shown in Table~\ref{table:detection}, ALLM-based baselines outperform SSL-based methods in cross-dataset ACC and F1. This pattern reflects the distribution shifts in the evaluation data. ADD involves noise addition and format conversion, while SF contains high-quality synthetic speech with multiple partially forged segments. Such variations undermine acoustic features specialized to training-distribution artifacts, limiting the generalization of SSL-based methods.

In contrast, ThinkOmni outperforms both SSL- and ALLM-based baselines, achieving absolute gains of 34.02\% in ACC and 34.74\% in F1 over the best SSL-based method (W2V2-Conformer), and gains of 4.57\% in ACC and 7.39\% in F1 over the best ALLM-based method (Qwen2-Audio), demonstrating strong cross-dataset generalization.

\subsection{Localization Evaluation}
Temporal manipulation localization is more challenging than spoofing detection due to the need for precise boundary prediction, especially in cross-dataset settings.
As shown in Table~\ref{table:localization}, we report mAP results of ThinkOmni alongside state-of-the-art SSL- and ALLM-based methods under both intra- and cross-dataset settings.

\noindent
\textbf{Intra-dataset Performance.}
The best SSL-based method, BAM, achieves 79.94\% mAP, while the strongest ALLM-based baseline, Qwen2.5-Omni-7B, reaches 83.69\%, indicating only a modest advantage of ALLMs in temporal localization. In contrast, ThinkOmni achieves the best performance with 88.05\% mAP, demonstrating the effectiveness of integrating multi-level forensic cues. Nevertheless, no method exceeds 90\% average mAP, underscoring the inherent difficulty of precise temporal manipulation localization.

\noindent
\textbf{Cross-dataset Performance.}
Cross-dataset temporal localization remains particularly challenging on SF, where a single utterance may contain multiple manipulated segments generated by different systems. SSL-based methods generalize poorly to such complex forgeries, with MRM and BAM achieving only 0.05\% and 4.40\% mAP, respectively. ALLM-based methods predict boundaries as discrete text tokens, which may limit the precision of multi-segment localization on unseen data; Qwen2.5-Omni-3B reaches only 12.08\% mAP on SF.
In contrast, ThinkOmni achieves a cross-dataset mAP of 74.67\%, exceeding the best SSL-based method (TDL) and the best ALLM-based method (Qwen2-Audio) by absolute margins of 43.73\% and 15.32\%, respectively, demonstrating strong generalization and precise localization on unseen data.

Overall, ThinkOmni outperforms SSL- and ALLM-based methods, achieving the best mAP and strong cross-dataset generalization, while baselines degrade under distribution shifts.

\begin{table}[t]
	\centering
	\caption{Step-wise Ablation of FACoT Construction in SFA, with ThinkOmni as reference.}
	\label{tab:Abla_FACoT}
	\begin{tabular*}{\columnwidth}{
			@{\extracolsep{\fill}}lcccccc@{}}
		\toprule
		\raisebox{-2.5ex}[0pt][0pt]{\textbf{Method}}
		& \multicolumn{3}{c}{\textbf{Intra-Dataset}}
		& \multicolumn{3}{c}{\textbf{Cross-Dataset}} \\
		\cmidrule(lr){2-4}
		\cmidrule(lr){5-7}
		& mACC & mF1 & mAP
		& mACC & mF1 & mAP \\
		\midrule
		Base data
		& 81.59 & 81.21 & 83.69
		& \underline{68.61} & \underline{75.23} & 55.91 \\
		
		\quad + CoT
		& 87.81 & 87.51 & 83.38
		& 64.99 & 67.69 & 60.27 \\
		
		\quad + CLAP
		& \underline{90.25}
		& \underline{90.42}
		& \underline{84.58}
		& 67.18
		& 72.87
		& \underline{63.35} \\
		\midrule
		ThinkOmni 
		& \textbf{93.70}
		& \textbf{93.72}
		& \textbf{88.05}
		& \textbf{80.74}
		& \textbf{85.15}
		& \textbf{74.67} \\
		\bottomrule
	\end{tabular*}
\end{table}

\subsection{Ablation Study}
This section presents ablation studies on FACoT construction, training strategies, and reasoning quality. Performance is evaluated using mean accuracy (mACC), mean F1 score (mF1), and mean average precision (mAP) under intra- and cross-dataset settings.

\noindent
\textbf{Ablation of FACoT Dataset.}
As shown in Table~\ref{tab:Abla_FACoT}, introducing CoT supervision improves intra-dataset detection and raises cross-dataset mAP from 55.91\% to 60.27\%, while reducing cross-dataset mACC and mF1. This trade-off suggests that reasoning supervision strengthens temporal evidence modeling and localization, but does not uniformly improve utterance-level generalization. CLAP-based filtering increases cross-dataset mAP to 63.35\%, demonstrating the importance of filtering weakly grounded rationale components for temporal localization. ThinkOmni achieves the best overall performance, confirming the complementary gains of FACoT supervision and the proposed training strategies.

\noindent
\textbf{Ablation of Learning Strategy in FMIL.}
Table~\ref{tab:Abla_FMIL} compares FMIL components, fusion methods, acoustic encoders, and training schedules. MFR alone performs poorly, while SAFE clearly outperforms naive concatenation with XLSR-300M. Among the evaluated acoustic encoders, XLSR-300M achieves the best performance.
Compared with joint training, progressive FMIL improves cross-dataset mACC, mF1, and mAP by 9.52\%, 9.99\%, and 3.95\%, respectively, achieving the best overall performance.

\noindent
\textbf{Ablation of Loss Components in FCML.}
Table~\ref{tab:Abla_FCML} presents the ablation of FCML loss components in the SFA stage. Replacing standard CE with weighted CE improves all metrics, including a 5.10\% gain in cross-dataset mAP. Adding $\mathcal{L}_{loc}$ further increases intra- and cross-dataset mAP by 2.22\% and 2.37\%, respectively. The complete FCML objective performs best, reaching 93.52\% mACC and 87.79\% mAP intra-dataset, and 69.79\% mACC and 70.82\% mAP cross-dataset. These results indicate that the largest advantage of ThinkOmni lies in preserving localization performance under distribution shift rather than merely improving in-domain accuracy.

\begin{table}[t]
	\centering
	\caption{Ablations of FMIL components, SAFE, acoustic encoders, and training schedules.}
	\label{tab:Abla_FMIL}
	\resizebox{\columnwidth}{!}{%
		\begin{tabular}{l ccc ccc}
			\toprule
			\raisebox{-2.5ex}[0pt][0pt]{\textbf{Method}} &
			\multicolumn{3}{c}{\textbf{Intra-Dataset}} &
			\multicolumn{3}{c}{\textbf{Cross-Dataset}} \\
			\cmidrule(lr){2-4}
			\cmidrule(lr){5-7}
			& mACC & mF1 & mAP & mACC & mF1 & mAP \\
			\midrule
			SFA
			& 93.52 & 93.63 & 87.79
			& 69.79 & 75.64 & 70.82 \\
			
			MFR
			& 51.87 & 50.45 & 58.15
			& 23.25 & 26.81 & 26.42 \\
			
			SFA+AFA (Concat, XLSR-300M)
			& 67.47 & 69.31 & 81.63
			& 56.33 & 67.95 & 70.21 \\
			
			SFA+AFA (SAFE, XLSR-300M)
			& \textbf{93.88} & \textbf{93.96} & 87.79
			& \underline{74.55} & \underline{80.34} & \underline{72.26} \\
			
			SFA+AFA (SAFE, XLSR-1B)
			& 85.05 & 84.90 & 81.13
			& 67.83 & 71.79 & 55.49 \\
			
			SFA+AFA (SAFE, BERT)
			& 89.37 & 89.36 & 84.10
			& 68.69 & 73.15 & 63.16 \\
			
			SFA+MFR
			& 93.66 & 93.75 & \textbf{88.35}
			& 68.19 & 73.88 & 69.01 \\
			
			\midrule
			Joint training
			& 93.67 & \underline{93.88} & \underline{88.24}
			& 71.22 & 75.16 & 70.72 \\
			
			ThinkOmni (ours)
			& \underline{93.70} & 93.72 & 88.05
			& \textbf{80.74} & \textbf{85.15} & \textbf{74.67} \\
			\bottomrule
		\end{tabular}%
	}
\end{table}

\begin{table}[t]
	\centering
	\caption{Ablation study of FCML loss configurations. Standard CE denotes the unweighted cross-entropy baseline.}
	\label{tab:Abla_FCML}
	\resizebox{\columnwidth}{!}{
		\begin{tabular}{l ccc ccc}
			\toprule
			\raisebox{-2.5ex}[0pt][0pt]{\textbf{Method}} &
			\multicolumn{3}{c}{\textbf{Intra-Dataset}} &
			\multicolumn{3}{c}{\textbf{Cross-Dataset}} \\
			\cmidrule(lr){2-4} \cmidrule(lr){5-7}
			& mACC & mF1 & mAP & mACC & mF1 & mAP \\
			\midrule
			Standard CE & 90.25 & 90.42 & 84.58 & 67.18 & 72.87 & 63.35 \\
			$\mathcal{L}_{wce}$ & \underline{91.84} & \underline{91.95} & \underline{85.57} & \underline{69.11} & \underline{74.44} & \underline{68.45} \\
			$\mathcal{L}_{wce}+\mathcal{L}_{loc}$ &
			\textbf{93.52} & \textbf{93.63} & \textbf{87.79} &
			\textbf{69.79} & \textbf{75.64} & \textbf{70.82} \\
			\bottomrule
	\end{tabular}}
\end{table}

\begin{figure}[t]
	\centering
	\includegraphics[width=\linewidth]{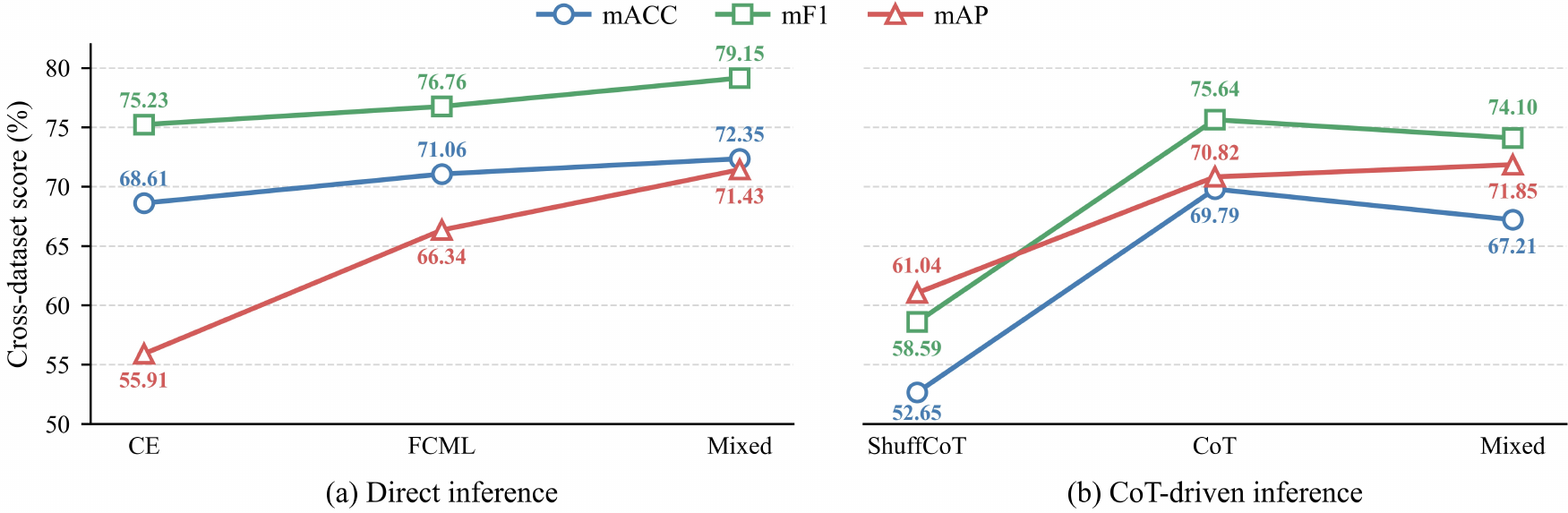}
	\caption{Ablation of CoT reasoning. The Mixed variants share the same checkpoint but differ in inference mode.}
	\label{fig:Abla_CoT}
\end{figure}

\noindent
\textbf{Ablation of Reasoning Capabilities.}
Figure~\ref{fig:Abla_CoT} shows the effects of reasoning supervision, inference-time CoT generation, and FMIL. Under direct inference, FCML improves mAP by 10.43\%, while mixed CoT/Non-CoT supervision further improves all metrics without rationale generation, indicating that CoT provides effective training guidance. For the same Mixed checkpoint, CoT-driven inference increases mAP by 0.42\% but decreases mACC and mF1 by 5.14\% and 5.05\%, respectively, revealing a detection-localization trade-off. ShuffCoT degrades all metrics, confirming the importance of sample-specific rationales. Finally, ThinkOmni surpasses the SFA-stage Mixed model by 13.53\%, 11.05\%, and 2.82\% in mACC, mF1, and mAP, respectively, confirming the benefits of FMIL.

\noindent
\textbf{Ablation of Reasoning Quality.}
We evaluate cross-dataset reasoning quality using automatic text metrics, MLLM judges, and human ratings, as reported in Table~\ref{tab:cot_quality}.

\textbf{1) Traditional Evaluation.}
ShuffCoT obtains the lowest BLEU-4 and CSS scores of 0.1247 and 0.6313, respectively, indicating limited agreement between mismatched rationales and the reference reasoning. In contrast, CoT, Mixed, and ThinkOmni achieve similar BLEU-4 scores (0.3058-0.3085) and CSS values
(0.8340-0.8357), showing high semantic similarity despite limited lexical overlap.

\textbf{2) MLLM and Human Evaluation.}
We sample 200 instances per class, yielding 600 samples, and evaluate each response on a 1--5 scale using three MLLM judges and twelve forensic researchers. 
The aggregate scores from each MLLM judge and the human evaluation yield the same ranking: Mixed, ThinkOmni, CoT, and ShuffCoT.
Mixed achieves the highest MLLM and human ratings, while ThinkOmni obtains comparable text-metric scores but slightly lower subjective ratings. The lower scores of ShuffCoT support the importance of sample-rationale alignment.
Evaluation details are provided in the supplementary material.

\begin{table}[t]
	\centering
	\caption{Reasoning quality across model variants. MLLM-judge and human-evaluation scores are rated on a 1--5 scale.}
	\label{tab:cot_quality}
	\resizebox{\columnwidth}{!}{
		\begin{tabular}{l cccc ccc c}
			\toprule
			\raisebox{-2.5ex}[0pt][0pt]{\textbf{Method}}
			& \multicolumn{4}{c}{\textbf{Text Metrics}}
			& \multicolumn{3}{c}{\textbf{MLLM Judges}}
			& \multicolumn{1}{c}{\textbf{Human}} \\
			\cmidrule(lr){2-5}
			\cmidrule(lr){6-8}
			\cmidrule(lr){9-9}
			& ROUGE-L & BLEU-4 & METEOR & CSS
			& Qwen & GPT & MiMo & Score \\
			\midrule
			
			ShuffCoT
			& 0.2646 & 0.1247 & 0.3729 & 0.6313
			& 3.2767 & 3.7167 & 3.5133 & 3.3933 \\
			
			CoT
			& 0.4599 & 0.3058 & 0.5657 & 0.8340
			& 4.3067 & 4.4000 & 4.1783 & 4.0467 \\
			
			Mixed
			& \underline{0.4612}
			& \textbf{0.3085}
			& \textbf{0.5708}
			& \textbf{0.8357}
			& \textbf{4.3883}
			& \textbf{4.5167}
			& \textbf{4.4333}
			& \textbf{4.2833} \\
			
			\textbf{ThinkOmni}
			& \textbf{0.4640}
			& \underline{0.3084}
			& \underline{0.5701}
			& \underline{0.8353}
			& \underline{4.3317}
			& \underline{4.4200}
			& \underline{4.3433}
			& \underline{4.1667} \\
			
			\bottomrule
		\end{tabular}}
\end{table}

\section{Conclusion}
We propose ThinkOmni, a reasoning-driven omni-modal LLM framework that shifts AFDL from implicit prediction toward evidence-driven analysis.
FACoT supplies structured, sample-aligned reasoning supervision; FMIL progressively integrates semantic, acoustic, and spectral-visual evidence; and FCML coordinates reasoning generation with detection and continuous boundary optimization.
Across intra- and cross-dataset evaluations, ThinkOmni consistently improves spoofing detection and temporal manipulation localization over the compared SSL- and ALLM-based methods.
Future work will pursue more efficient inference, calibrated abstention, broader robustness evaluation, and specialized audio-forensic reward models for more faithful fine-grained reasoning.

\section{Acknowledgements}
This work was supported in part by the National Natural Science Foundation of China (NSFC) under Grant U23B2022; in part by the Guangdong Basic and Applied Basic Research Foundation under Grant 2025A1515010234; in part by Shenzhen Science and Technology Program under Grant SYSPG20241211174032004 and JCYJ20250604181211016.

\putbib[ref]
\end{bibunit}

\clearpage
\appendix
\begin{bibunit}[ACM-Reference-Format]
\makeatletter
\def\@extra@binfo{@supp}
\def\@extra@b@citeb{@supp}
\makeatother

\renewcommand{\shortauthors}{ThinkOmni: Supplementary Materials}
\renewcommand{\shorttitle}{ThinkOmni: Supplementary Materials}
\twocolumn[
\begin{center}
	{\Huge\bfseries ThinkOmni: A Reasoning-Driven Omni-Modal LLM Framework\par}
	\vspace{0.25em}
	{\Huge\bfseries for Audio Forgery Detection and Localization\par}
	\vspace{0.65em}
	{\Large\bfseries (Supplementary Materials)\par}
	\vspace{1em}
\end{center}
]
\sloppy

\definecolor{mycol2}{HTML}{2EC4B6}
\definecolor{promptorange}{RGB}{255, 153, 51} 

\newtcolorbox{promptbox}{
	breakable,
	colback=mycol2!8!white,
	colframe=mycol2!65!black,
	coltitle=black,
	arc=6pt,
	boxrule=0.8pt,
	left=10pt,
	right=10pt,
	top=10pt,
	bottom=10pt,
	upperbox=visible
}

\newtcolorbox{mytocbox}[1][]{
	colback=mycol2!8!white,
	colframe=mycol2!65!white,
	title={\textbf{Appendix Contents}},
	coltitle=black,
	arc=2mm,
	boxrule=1pt,
	#1
}

\begin{mytocbox}
	\setlength{\parskip}{0.4em} 
	\noindent
	\hypersetup{linkcolor=black!60!black}
	
	\textbf{A.} \hyperref[supp:CoTReasoning]{Implicit vs. Chain-of-Thought Reasoning} \dotfill P\pageref{supp:CoTReasoning} 
	
	\textbf{B.} \hyperref[supp:framework]{Details of ThinkOmni Framework} \dotfill P\pageref{supp:framework} 
	
	\hspace*{1.5em} \textbf{B.1} \hyperref[supp:framework_afe]{Acoustic Feature Extraction} \dotfill P\pageref{supp:framework_afe} 
	
	\hspace*{1.5em} \textbf{B.2} \hyperref[supp:framework_safe]{Semantic-Acoustic Forensic Enhancer} \dotfill P\pageref{supp:framework_safe} 
	
	\textbf{C.} \hyperref[supp:dataset]{Details of FACoT Dataset} \dotfill P\pageref{supp:dataset}
	
	\hspace*{1.5em} \textbf{C.1} \hyperref[supp:dataset_test_id]{Intra-Dataset Evaluation} \dotfill P\pageref{supp:dataset_test_id} 
	
	\hspace*{1.5em} \textbf{C.2} \hyperref[supp:dataset_test_cd]{Cross-Dataset Evaluation} \dotfill P\pageref{supp:dataset_test_cd} 
	
	\hspace*{1.5em} \textbf{C.3} \hyperref[supp:data_plat]{Data Correction Platform} \dotfill P\pageref{supp:data_plat} 
	
	\hspace*{1.5em} \textbf{C.4} \hyperref[supp:annotation_quality]{FACoT Annotation Pipeline} \dotfill P\pageref{supp:annotation_quality}
	
	\hspace*{1.5em} \textbf{C.5} \hyperref[supp:data_stat]{FACoT Dataset Statistics} \dotfill P\pageref{supp:data_stat} 
	
	\textbf{D.} \hyperref[supp:baseline_method]{Details of Baseline Methods} \dotfill P\pageref{supp:baseline_method}
	
	\textbf{E.} \hyperref[supp:more_imple]{More Implementation Details} \dotfill P\pageref{supp:more_imple}
	
	\hspace*{1.5em} \textbf{E.1} \hyperref[supp:data_prep]{Data Preprocessing}  \dotfill P\pageref{supp:data_prep} 
	
	\hspace*{1.5em} \textbf{E.2} \hyperref[supp:model_conf]{Model Configuration} \dotfill P\pageref{supp:model_conf} 
	
	\hspace*{1.5em} \textbf{E.3} \hyperref[supp:train_stra]{Detailed Training Strategy} \dotfill P\pageref{supp:train_stra} 
	
	\hspace*{1.5em} \textbf{E.4} \hyperref[supp:inf_conf]{Inference Configuration} \dotfill P\pageref{supp:inf_conf} 
	
	\textbf{F.} \hyperref[supp:more_results]{More Experimental Results} \dotfill P\pageref{supp:more_results}
	
	\hspace*{1.5em} \textbf{F.1} \hyperref[supp:det_results]{Detection Results}  \dotfill P\pageref{supp:det_results} 
	
	\hspace*{1.5em} \textbf{F.2} \hyperref[supp:loc_results]{Localization Results} \dotfill P\pageref{supp:loc_results} 
	
	\hspace*{1.5em} \textbf{F.3} \hyperref[supp:token_results]{Effect of Token Weighting} \dotfill P\pageref{supp:token_results} 
	
	\hspace*{1.5em} \textbf{F.4} \hyperref[supp:computational_efficiency]{Computational Efficiency} \dotfill P\pageref{supp:computational_efficiency}
	
	\hspace*{1.5em} \textbf{F.5} \hyperref[supp:cot_stage_ablation]{Stage-wise Effect of CoT} \dotfill P\pageref{supp:cot_stage_ablation}
	
	\hspace*{1.5em} \textbf{F.6} \hyperref[supp:safe_ablation]{Effect of SAFE Fusion} \dotfill P\pageref{supp:safe_ablation}
	
	\textbf{G.} \hyperref[supp:case_study]{Case Study} \dotfill P\pageref{supp:case_study}
	
	\hspace*{1.5em} \textbf{G.1} \hyperref[supp:success_cases]{Successful Case Studies}  \dotfill P\pageref{supp:success_cases} 
	
	\hspace*{1.5em} \textbf{G.2} \hyperref[supp:failed_cases]{Failure Case Studies} \dotfill P\pageref{supp:failed_cases} 
	
	\textbf{H.} \hyperref[supp:full_prompts]{Prompt Templates} \dotfill P\pageref{supp:full_prompts}
	
	\hspace*{1.5em} \textbf{H.1} \hyperref[supp:annotation_system_prompt]{FACoT System Prompt} \dotfill P\pageref{supp:annotation_system_prompt}
	
	\hspace*{1.5em} \textbf{H.2} \hyperref[supp:annotation_user_prompt]{FACoT User Prompt} \dotfill P\pageref{supp:annotation_user_prompt}
	
	\hspace*{1.5em} \textbf{H.3} \hyperref[supp:thinkomni_input_prompt]{ThinkOmni Input Prompt} \dotfill P\pageref{supp:thinkomni_input_prompt}
	
	\hspace*{1.5em} \textbf{H.4} \hyperref[supp:mllm_quality_prompt]{MLLM Evaluation Prompt} \dotfill P\pageref{supp:mllm_quality_prompt}
\end{mytocbox}

\section{Implicit vs. Chain-of-Thought Reasoning}
\label{supp:CoTReasoning}
In this section, we compare two reasoning paradigms in audio large language model (ALLM)-based audio forensics: implicit reasoning and chain-of-thought (CoT) reasoning. Our goal is to characterize their mechanisms by defining the input and output variables, along with their probabilistic factorization forms.

Let the input be $\boldsymbol{X}=(\boldsymbol{A},\boldsymbol{S},\boldsymbol{I})$, where $\boldsymbol{A}$ is the raw audio waveform, $\boldsymbol{S}$ is its spectrogram, and $\boldsymbol{I}$ is the forensic instruction. ThinkOmni generates $\boldsymbol{Y}=(\boldsymbol{r},c,\boldsymbol{z})$, where $\boldsymbol{r}=(r_1,\ldots,r_{N_r})$ is the forensic reasoning sequence, $c\in\{0,1,2\}$ denotes fully real, fully fake, and partially fake audio, respectively, and $\boldsymbol{z}$ is the timestamp-token sequence. The parser $g(\cdot)$ converts $\boldsymbol{z}$ into a set of zero or more temporal intervals, $\mathcal{B}=g(\boldsymbol{z})=\{[s_k,e_k]\}_{k=1}^{K}$. This definition supports multiple manipulated segments and is consistent with the task formulation in the main paper.

\noindent\textbf{Implicit Reasoning.}
In implicit reasoning \cite{Li2025}, the model does not explicitly generate the reasoning sequence $\boldsymbol{r}$ and directly predicts the detection label $c$ and timestamp-token sequence $\boldsymbol{z}$ from $\boldsymbol{X}$. The conditional probability is formulated as:
\begin{equation}
	P_{\theta}(c,\boldsymbol{z}\mid\boldsymbol{X}),
\end{equation}
which corresponds to a single-stage mapping $f:\boldsymbol{X}\rightarrow(c,\boldsymbol{z})$. In this formulation, $\boldsymbol{r}$ is not part of the output space, and any intermediate inference remains latent in the model parameters.

\noindent\textbf{Chain-of-Thought (CoT) Reasoning.}
In contrast, CoT reasoning \cite{wei2022chain} treats $\boldsymbol{r}$ as an explicit component of $\boldsymbol{Y}$. The joint probability is factorized into a reasoning stage followed by detection and timestamp generation:
\begin{equation}
	P_{\theta}(\boldsymbol{r},c,\boldsymbol{z}\mid\boldsymbol{X})
	=
	\prod_{t=1}^{N_r}P_{\theta}(r_t\mid\boldsymbol{X},r_{<t})
	\cdot
	P_{\theta}(c,\boldsymbol{z}\mid\boldsymbol{X},\boldsymbol{r}).
\end{equation}
Accordingly, the mapping is decomposed as $f':\boldsymbol{X}\rightarrow\boldsymbol{r}\rightarrow(c,\boldsymbol{z})$, where detection and timestamp generation are conditioned on both the multi-modal input and the explicit reasoning sequence.

The key difference is whether $\boldsymbol{r}$ is explicitly supervised and generated. CoT factorizes the task into rationale generation and target prediction, encouraging the model to organize forensic cues before producing the detection label and timestamp sequence. This formulation provides an explicit intermediate supervision signal; its empirical effect is evaluated through the reasoning ablations rather than assumed from the factorization alone.

\section{Details of ThinkOmni Framework}
\label{supp:framework}
In this part, we provide detailed architectural descriptions and mathematical formulations for the core components of the \textbf{ThinkOmni} framework. Specifically, we detail the acoustic feature extraction module and the Semantic-Acoustic Forensic Enhancer (SAFE) module, which performs dual-branch (local and global) feature fusion.

\subsection{Acoustic Feature Extraction}
\label{supp:framework_afe}
To capture fine-grained acoustic artifacts and low-level spoofing traces, ThinkOmni utilizes a pre-trained Wav2Vec 2.0 XLSR-300M model\footnote{https://huggingface.co/facebook/wav2vec2-xls-r-300m}. 
Instead of solely relying on the final layer's output, we leverage the hierarchical representations learned across different depths of the network.

Given the input audio waveform, the acoustic encoder extracts hidden states from the last $L=24$ Transformer layers. 
Let $\boldsymbol{H}_l \in \mathbb{R}^{T_{xlsr} \times D_{xlsr}}$ denote the hidden state from the $l$-th layer. We compute the final acoustic representation $\boldsymbol{F}_{xlsr}$ as a dynamically weighted sum of these layers:
\begin{equation}
	\boldsymbol{F}_{xlsr} = \sum_{l=1}^{L} \alpha_l \boldsymbol{H}_l, \quad \text{where} \quad \alpha_l = \frac{\exp(w_l)}{\sum_{j=1}^{L} \exp(w_j)}
\end{equation}
where $\{w_1, \dots, w_L\}$ are learnable parameters. This layer-wise aggregation allows the model to adaptively focus on the specific feature levels that are most indicative of audio forgery.

\subsection{Semantic-Acoustic Forensic Enhancer}
\label{supp:framework_safe}
The core of our cross-modal alignment is the SAFE module, which integrates semantic features $\boldsymbol{F}_{sem} \in \mathbb{R}^{T \times D_{sem}}$ and acoustic features $\boldsymbol{F}_{xlsr} \in \mathbb{R}^{T_{xlsr} \times D_{xlsr}}$. The SAFE module consists of forgery-aware positional encoding, a local cross-attention branch, a global forgery discriminator, and a gated fusion mechanism.

\noindent\textbf{Forgery-Aware Positional Encoding.}
To preserve sequential structure before cross-modal fusion, we add scaled sinusoidal positional encodings to the semantic features. A frequency scaling factor of $s=1.5$ is introduced to better capture the temporal patterns of forgery artifacts. The positional encoding is defined as:
\begin{equation}
	\begin{aligned}
		\boldsymbol{E}_{(pos, 2i)} &= \sin\left(\frac{pos}{10000^{2i/D_{sem}}} \cdot s \right), \\
		\boldsymbol{E}_{(pos, 2i+1)} &= \cos\left(\frac{pos}{10000^{2i/D_{sem}}} \cdot s \right).
	\end{aligned}
\end{equation}

The position-enhanced semantic features are computed as $\tilde{\boldsymbol{F}}_{sem} = \boldsymbol{F}_{sem} + \boldsymbol{E}$.

\noindent\textbf{Local Cross-Attention.}
To capture fine-grained alignment between semantic content and acoustic anomalies, we employ local cross-attention. The acoustic features $\boldsymbol{F}_{xlsr}$ are first projected and temporally interpolated to match the semantic sequence length $T$, yielding $\widetilde{\boldsymbol{F}}_{xlsr}\in\mathbb{R}^{T\times 
	D_{sem}}$.

To suppress redundant acoustic variations and reduce computational overhead, we project both modalities into a shared low-rank bottleneck space with dimension $D_k$:
\begin{equation}
	\boldsymbol{Q} = \tilde{\boldsymbol{F}}_{sem} \boldsymbol{W}_q, \quad \boldsymbol{K} = \tilde{\boldsymbol{F}}_{xlsr} \boldsymbol{W}_k, \quad \boldsymbol{V} = \tilde{\boldsymbol{F}}_{xlsr} \boldsymbol{W}_v,
\end{equation}
where $\boldsymbol{W}_q,\boldsymbol{W}_k,\boldsymbol{W}_v\in\mathbb{R}^{D_{sem}\times D_k}$ are learnable projection matrices. The local fused representation is then obtained through scaled dot-product attention with a residual connection:
\begin{equation}
	\boldsymbol{F}_{local} = \tilde{\boldsymbol{F}}_{sem} + \text{Softmax}\left(\frac{\boldsymbol{Q} \boldsymbol{K}^\top}{\sqrt{D_k}}\right) \boldsymbol{V} \boldsymbol{W}_{out},
	\label{eq:loc}
\end{equation}
where $\boldsymbol{W}_{out} \in \mathbb{R}^{D_k \times D_{sem}}$.
Equation~(\ref{eq:loc}) lets each semantic position attend to the temporally aligned acoustic sequence and then adds the attended acoustic feature through a residual connection. This branch is designed to expose token-level semantic--acoustic interactions that may assist temporal boundary prediction.

\noindent\textbf{Global Forgery Discriminator.}
While the local branch focuses on frame-level alignment, the global branch captures long-range spoofing patterns and holistic inconsistencies. Specifically, the global forgery discriminator extracts sequence-level representations by applying length-aware mean pooling to both modalities, enabling robust aggregation of temporal information while accounting for variable input durations.

Let $\bar{\boldsymbol{F}}_{sem} \in \mathbb{R}^{D_{sem}}$ and $\bar{\boldsymbol{F}}_{xlsr} \in \mathbb{R}^{D_{xlsr}}$ be the temporally pooled features. We map them to the same latent space using multilayer perceptrons (MLPs), each consisting of a Linear-LayerNorm-GELU-Dropout-Linear sequence:
\begin{equation}
	\boldsymbol{Z}_{sem} = \text{MLP}_{sem}(\bar{\boldsymbol{F}}_{sem}), \quad \boldsymbol{Z}_{xlsr} = \text{MLP}_{xlsr}(\bar{\boldsymbol{F}}_{xlsr}).
\end{equation}

The global forgery feature $\boldsymbol{F}_{global}\in\mathbb{R}^{D_{sem}}$ is obtained by concatenating the two representations and passing them through a fusion block:
\begin{equation}
	\boldsymbol{F}_{global} =
	\text{Dropout}\left(
	\text{GELU}\left(
	\text{LayerNorm}\left(
	\text{Linear}\left([\boldsymbol{Z}_{sem} \parallel \boldsymbol{Z}_{xlsr}]\right)
	\right)
	\right)
	\right),
\end{equation}
where $\parallel$ denotes the concatenation operation.

\noindent\textbf{Gated Multi-level Fusion.}
To selectively integrate local frame-level alignments and global sequence-level context, we employ a gated fusion mechanism. We first obtain the pooled local representation as $\bar{\boldsymbol{F}}_{local}=\operatorname{MeanPool}(\boldsymbol{F}_{local})$. 
A dynamic gate vector $\boldsymbol{G}\in\mathbb{R}^{D_{sem}}$ is then computed as:
\begin{equation}
	\boldsymbol{G} = \sigma\left( \boldsymbol{W}_{gate} [\bar{\boldsymbol{F}}_{local} \parallel \boldsymbol{F}_{global}] \right),
\end{equation}
where $\sigma$ denotes the sigmoid activation function and $\boldsymbol{W}_{gate} \in \mathbb{R}^{D_{sem} \times 2D_{sem}}$. The final fused output $\boldsymbol{F}_{out}$ is generated by modulating the global feature with the gate and adding it to the local feature, followed by normalization:
\begin{equation}
	\boldsymbol{F}_{out} = \text{Dropout}\left( \text{LayerNorm}\left( \boldsymbol{F}_{local} + \boldsymbol{G} \odot \boldsymbol{F}_{global} \right) \right),
\end{equation}
where $\odot$ denotes element-wise multiplication, and $\boldsymbol{G}\odot\boldsymbol{F}_{global}$ is broadcast along the temporal dimension. The fused sequence combines the token-level branch with a gated sequence-level feature before it is passed to the LLM reasoning backbone.

\begin{table}[t]
	\centering
	\caption{Class distributions of the training and test sets across datasets.}
	\label{tab:dataset_stats}
	\resizebox{\linewidth}{!}{
		\begin{tabular}{c | ccc | ccc}
			\toprule
			\raisebox{-4ex}[0pt][0pt]{Dataset} & \multicolumn{3}{c|}{Training Set} & \multicolumn{3}{c}{Test Set} \\ 
			\cmidrule(lr){2-4} \cmidrule(lr){5-7}
			&\begin{tabular}[c]{@{}c@{}}Fully \\ Real\end{tabular} & \begin{tabular}[c]{@{}c@{}}Fully \\ Fake\end{tabular} & \begin{tabular}[c]{@{}c@{}}Partially \\ Fake\end{tabular} 
			&\begin{tabular}[c]{@{}c@{}}Fully \\ Real\end{tabular} & \begin{tabular}[c]{@{}c@{}}Fully \\ Fake\end{tabular} & \begin{tabular}[c]{@{}c@{}}Partially \\ Fake\end{tabular}  \\ 
			\midrule
			PS        & 5,128 & 11,360 & 5,679 & 3,333 & 3,333 & 3,334 \\
			HAD       & 5,679 & 1,615  & 5,679 & 0     & 234   & 8,838 \\
			LAV-DF    & 5,679 & 0      & 5,679 & 5,000 & 0     & 5,000 \\
			SINE      & 5,679 & 5,679  & 5,679 & 3,334 & 3,333 & 3,333 \\
			LPS       & 2,391 & 5,679  & 5,679 & 3,333 & 3,334 & 3,333 \\
			ArEnAV    & 5,679 & 0      & 5,679 & 5,000 & 0     & 5,000 \\
			AV-1M++   & 5,679 & 0      & 5,679 & 5,000 & 0     & 5,000 \\
			ADD       & 0     & 0      & 0     & 5,000 & 0     & 5,000 \\
			SF        & 0     & 0      & 0     & 4,583 & 0     & 5,045 \\ 
			\midrule
			\textbf{Total} & \textbf{35,914} & \textbf{24,333} & \textbf{39,753} & \textbf{34,583} & \textbf{10,234} & \textbf{43,883} \\ 
			\bottomrule
	\end{tabular}}
\end{table}

\section{Details of FACoT Dataset}
\label{supp:dataset}
FACoT comprises 100K training samples aggregated from eight public benchmarks, covering diverse attacks, languages, and acoustic conditions. Intra-dataset evaluation uses non-overlapping test samples from the source benchmarks, while ADD and Speech-Forensics serve as external cross-dataset test sets. Table~\ref{tab:dataset_stats} summarizes the class distributions of the training and evaluation sets.

Although FACoT includes eight training sources, the intra-dataset evaluation contains seven groups. Since PartialSpoof is derived from ASVspoof 2019 LA, fully fake 19LA test samples are assigned to the PS group to avoid duplication. The remaining groups use non-overlapping test samples from their respective source datasets. For AV-Deepfake1M++, we use the development set because test labels are unavailable. ADD and Speech-Forensics are used exclusively for cross-dataset evaluation.

\subsection{Intra-Dataset Evaluation}
\label{supp:dataset_test_id}
We construct the training set from the following eight datasets, with intra-dataset evaluation splits derived accordingly:
\begin{itemize}
	\item \textbf{ASVspoof 2019 LA (19LA) \cite{ASV2019}:} A fundamental benchmark for Logical Access (LA) attacks, encompassing various Text-to-Speech (TTS) and Voice Conversion (VC) generated spoofing audio samples. Because PartialSpoof is derived from 19LA, the fully fake 19LA test samples are reported within the PS evaluation group, and overlapping source utterances are removed across the training and test splits.
	\item \textbf{PartialSpoof (PS) \cite{partialspoof}:} Derived from the 19LA dataset, this benchmark is the first English dataset for partially deepfake audio. It focuses on partial spoofing by concatenating real and fake segments, and provides fine-grained temporal boundaries, where segments are randomly replaced between genuine and spoofed audio, with both segment-level and utterance-level labels annotated based on the presence of spoofed content.
	\item \textbf{Half-Truth (HAD) \cite{HAD}:} The first Chinese dataset for partially deepfake audio, built on the AISHELL-3 corpus \cite{Shi2021}, comprising partially fake, fully fake, and real samples. Unlike the PS database, manipulations preserve semantic coherence and word boundaries rather than random segment replacement, and include precise start and end timestamps for forged intervals.
	\item \textbf{LAV-DF \cite{LAV-DF}:} The first content-driven audio-visual deepfake dataset for temporal manipulation localization, where manipulations alter semantic content (e.g., sentiment polarity) with fine-grained temporal annotations. In our audio-only setting, we use only the audio modality.
	\item \textbf{SINE \cite{SINE}:} A large-scale dataset for seamless partially deepfake audio, constructed using neural speech infilling models (e.g., Voicebox) to generate edits with smooth transitions, avoiding the discontinuities introduced by traditional cut-and-paste methods. It includes both authentic and edited speech with fine-grained temporal annotations, and is designed to support detection and localization of seamless speech manipulations.
	\item \textbf{LlamaPartialSpoof (LPS) \cite{Llamapartialspoof}:} A content-driven audio-only deepfake dataset built upon LibriTTS. It enhances the diversity of fully and partially fake utterances by using Llama-3-8B-Instruct to modify transcripts via prompts, producing more natural manipulations. Five TTS models generate the fake audio, with partially fake samples formed by concatenating real and synthesized segments, and post-processing applied to all utterances.
	\item \textbf{ArEnAV \cite{ArEnAV}:} A bilingual (Arabic and English) audio-visual deepfake dataset with intra-utterance code-switching and dialectal variation, containing large-scale real and fake videos generated via TTS and lip-sync models for multilingual deepfake detection.
	\item \textbf{AV-Deepfake1M++ (AV-1M++) \cite{Av-deepfake1m++}:} A large-scale audio-visual deepfake benchmark with over 2M clips, featuring diverse manipulation strategies and real-world perturbations, with fine-grained annotations for detection and temporal localization. As the test set labels are not publicly available, the development set is used for evaluation.
\end{itemize}

\subsection{Cross-Dataset Evaluation}
\label{supp:dataset_test_cd}
For cross-dataset evaluation, ADD and Speech-Forensics are reserved exclusively for testing, and no samples from either dataset are used during training.
\begin{itemize}
	\item \textbf{ADD 2023 Track 2 (ADD) \cite{yi2023add}:} It is designed for the second Audio Deep Synthesis Detection Challenge (ADD 2023) and includes fully fake, partially fake, and genuine audio. Partially fake samples are generated by replacing segments of authentic audio with either real or synthesized clips. The training and development sets contain all three types, whereas the test set features unseen partially fake and real utterances. Moreover, noise and format conversions are applied to the test data, substantially increasing the difficulty of manipulation localization.
	\item \textbf{Speech-Forensics (SF) \cite{Speech-Forensics}:} This dataset contains diverse audio manipulations with segment-level boundaries and synthesis-method labels. Its multi-segment and multi-system samples support evaluation of forgery detection and temporal localization under distribution shift.
\end{itemize}

\subsection{Data Correction Platform}
\label{supp:data_plat}
\begin{figure*}[t]
	\centering
	\begin{subfigure}[b]{0.33\linewidth}
		\centering
		\includegraphics[width=\linewidth]{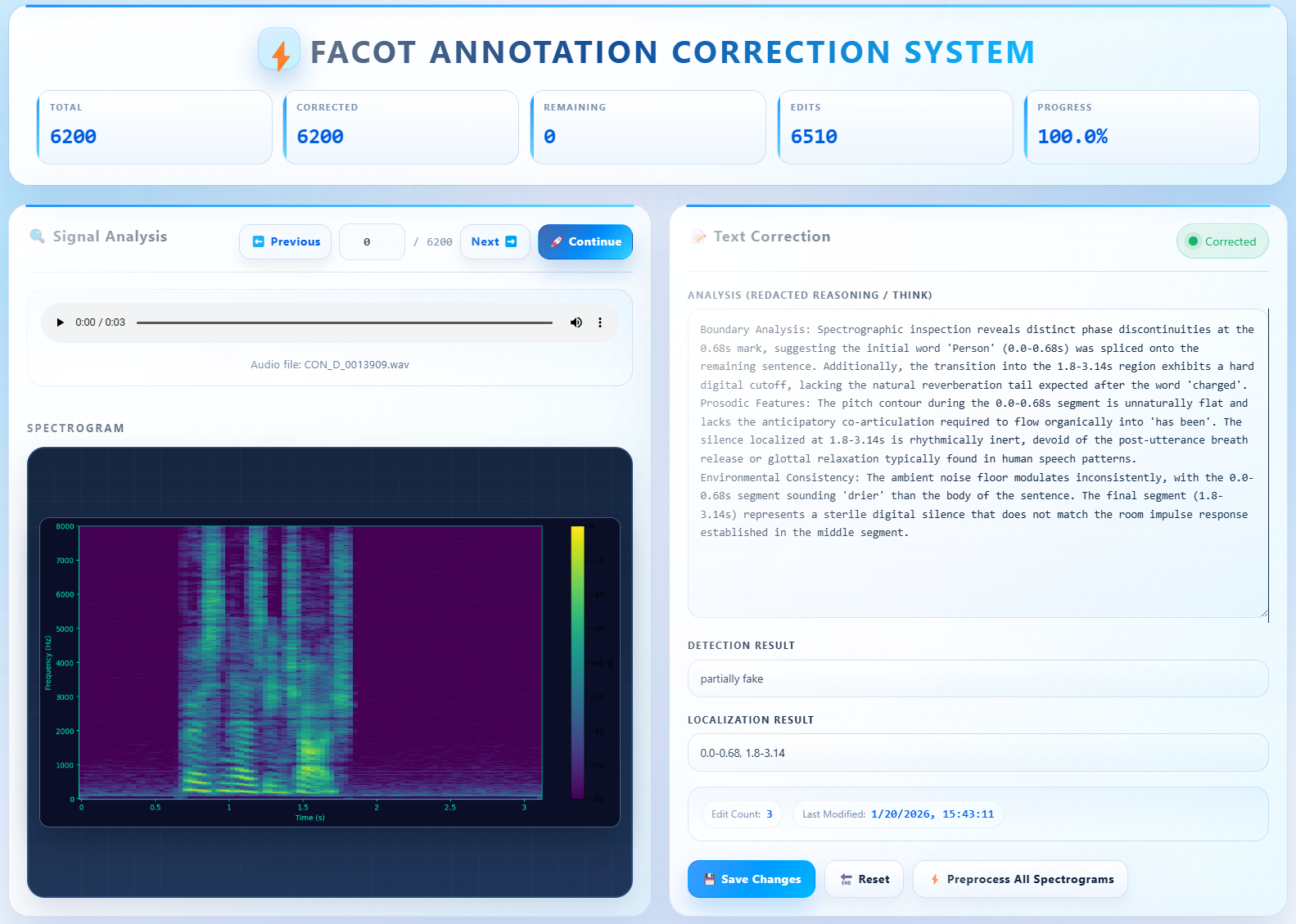}
		\caption{Correction platform interface}
		\label{fig:ann_system}
	\end{subfigure}
	\begin{subfigure}[b]{0.32\linewidth}
		\centering
		\includegraphics[width=\linewidth]{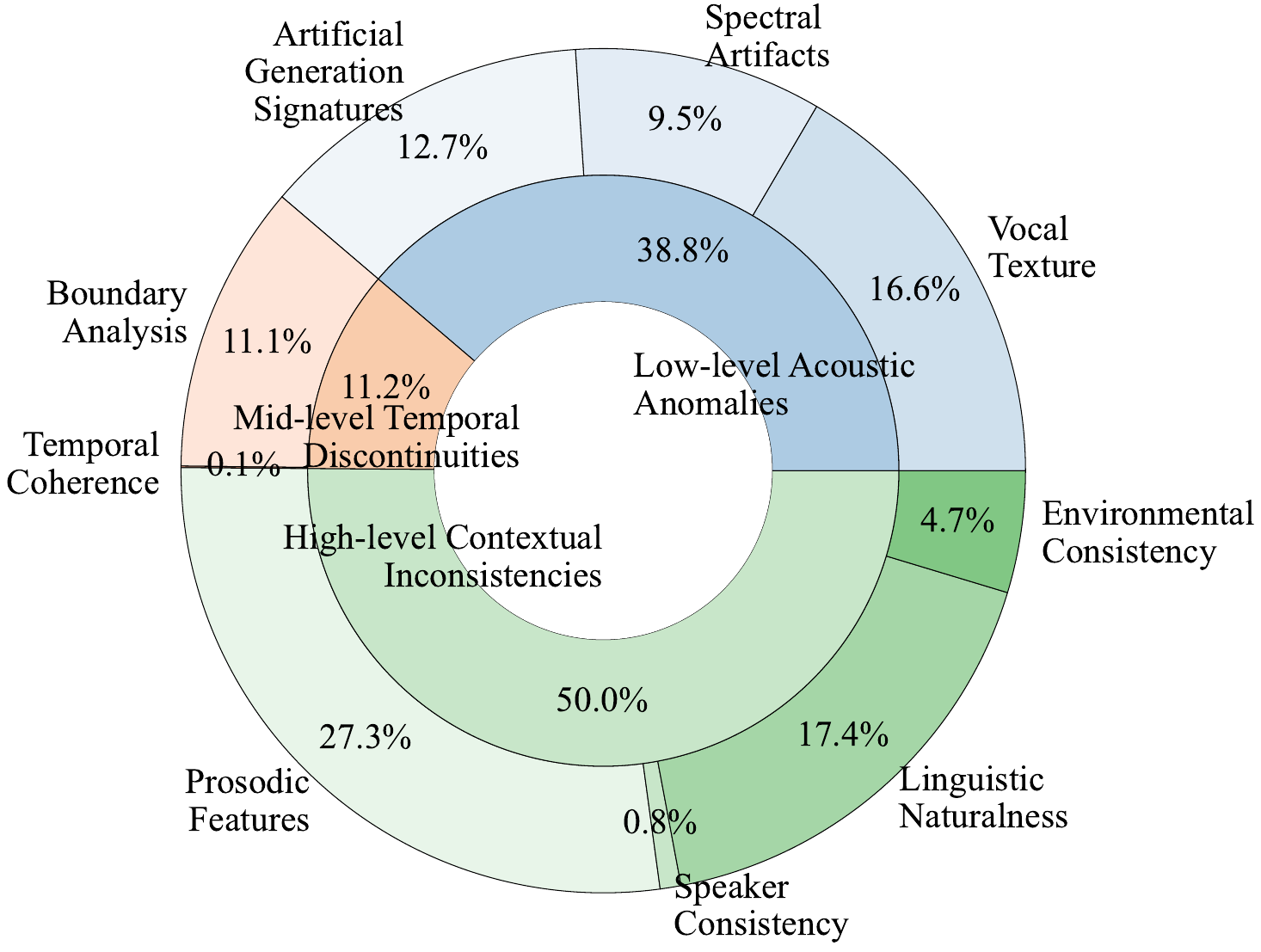}
		\caption{Distribution of annotation dimensions}
		\label{fig:ann_dist}
	\end{subfigure}
	\begin{subfigure}[b]{0.34\linewidth}
		\centering
		\includegraphics[width=\linewidth]{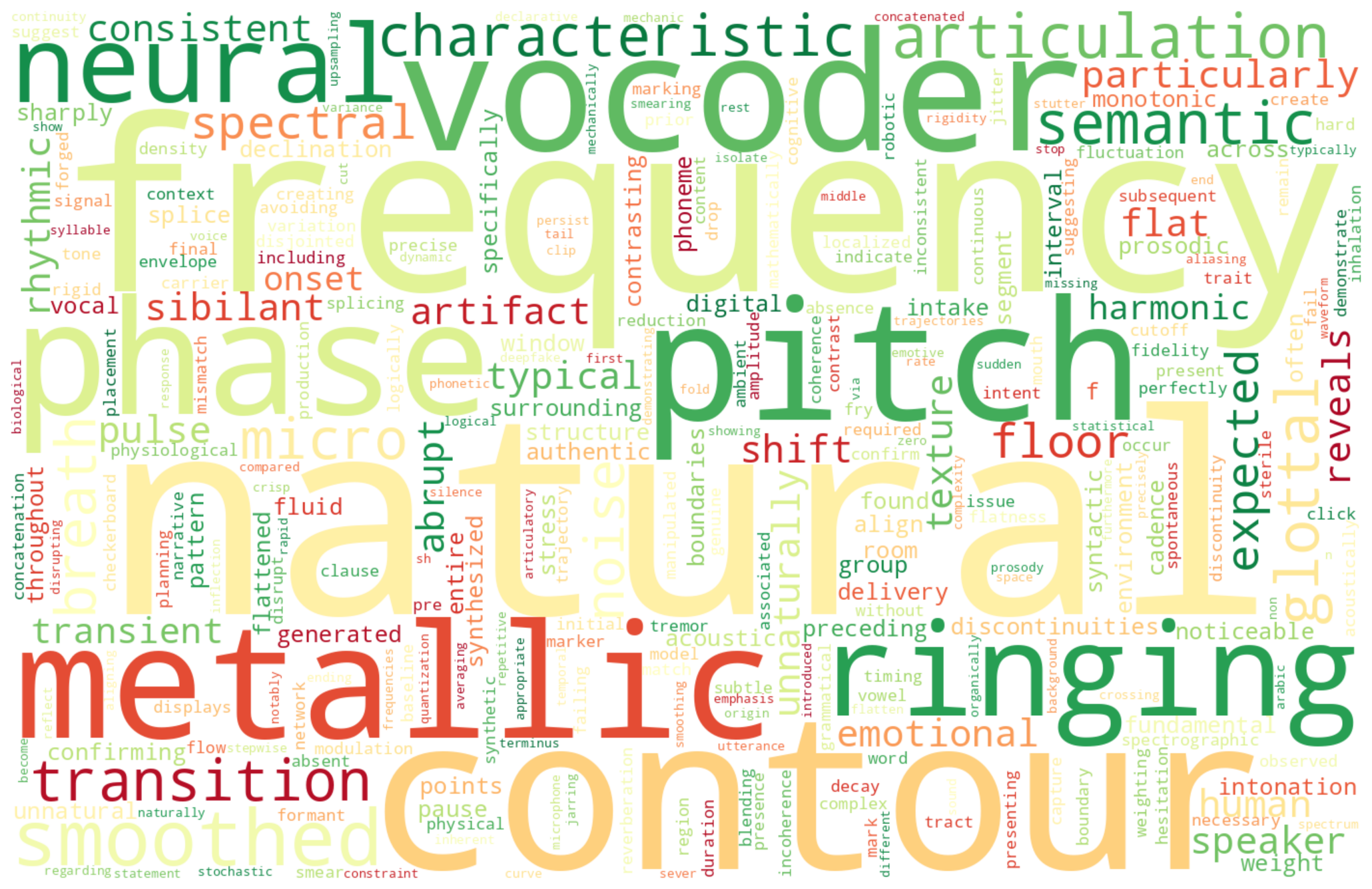}
		\caption{Word cloud of FACoT annotations}
		\label{fig:wordcloud}
	\end{subfigure}
	\caption{Overview of FACoT annotation correction and analysis: (a) correction platform interface, (b) distribution of annotation dimensions, and (c) word cloud of FACoT annotations.}
\end{figure*}

Figure~\ref{fig:ann_system} shows the annotation-correction interface, which combines waveform and spectrogram visualization with LLM-generated rationales. The interface supports expert review of predictions, acoustic evidence, timestamps, and annotation text; it is a review tool rather than independent evidence of annotation accuracy.

\subsection{FACoT Annotation Pipeline}
\label{supp:annotation_quality}
\noindent
\textbf{Annotation Protocol.}
FACoT adopts a label-aware annotation protocol. For each seed sample, the annotation model receives the audio, spectrogram, reference authenticity label, and temporal boundaries. The generated rationale must describe observable forensic evidence rather than merely restating the label or timestamps. Self-curation verifies its logical consistency with the reference metadata, while expert verification assesses transcript accuracy, localized evidence, timestamp alignment, acoustic continuity, speaker consistency, physiological plausibility, and frequency-level justification.

\noindent
\textbf{CoT Annotation.}
The 6.2K seed samples are selected through stratified sampling across source datasets and authenticity classes. Given the reference labels and temporal boundaries, Gemini-3-Pro \cite{gemini} generates a structured rationale for each sample. After self-curation, a forensic expert evaluates each annotation using the eight-item checklist described above. The verified seed set is then used to adapt Qwen3-Omni \cite{Qwen3-omni}, which generates annotations for the remaining 93.8K samples. Thus, expert verification establishes the annotation schema and quality standard through the seed set, while the remaining annotations are model-generated and filtered.

\noindent
\textbf{Semantic Quality Filtering.}
After large-scale annotation, CLAP filtering is applied to each reasoning dimension. Dimensions with audio--text similarity below 0.2 are removed, while the corresponding audio samples and remaining rationale components are retained. Since CLAP measures audio--text compatibility, it suppresses weakly grounded content but cannot certify causal faithfulness or validate every localized claim. We therefore use CLAP as a semantic quality filter rather than a substitute for expert forensic verification.

\subsection{FACoT Dataset Statistics}
\label{supp:data_stat}
Figure~\ref{fig:ann_dist} summarizes the retained FACoT annotation dimensions. High-level contextual dimensions account for 50.0\% of retained components, followed by low-level acoustic dimensions (38.8\%) and mid-level temporal dimensions (11.2\%). At finer granularity, prosodic features account for 27.3\%, linguistic naturalness for 17.4\%, vocal texture for 16.6\%, generation signatures for 12.7\%, boundary analysis for 11.1\%, spectral artifacts for 9.5\%, Environmental Consistency for 4.7\%, speaker consistency for 0.8\%, and temporal coherence for 0.1\%. The word cloud in Figure~\ref{fig:wordcloud} likewise shows frequent annotation terms such as ``pause,'' ``pitch,'' and ``boundary.''

\section{Details of Baseline Methods}
\label{supp:baseline_method}
We benchmark ThinkOmni against a comprehensive suite of state-of-the-art (SOTA) methods. These baselines are broadly categorized into traditional Self-Supervised Learning (SSL)-based methods and recent Audio Large Language Model (ALLM)-based methods.

\noindent
\textbf{Task Harmonization.}
All detection methods are evaluated in a three-class setting of fully real, fully fake, and partially fake audio. Following the main protocol, ThinkOmni and all baselines are trained on the same 100K-sample FACoT pool. Each baseline uses only the forgery labels or temporal boundaries required by its objective, while structured FACoT rationales are used only by ThinkOmni and its reasoning-based variants. Detection performance is measured using accuracy and F1 across the three classes.

For detection-only methods, partially fake utterances are treated as an independent class rather than merged with fully fake audio. Localization methods receive all reference intervals for partially fake samples, no interval for fully real samples, and the full-utterance interval for fully fake samples. This protocol maintains consistent training data and label semantics while preserving the original optimization objective of each baseline family.

\noindent\textbf{SSL-based Methods for Spoofing Detection.}
These methods map acoustic features directly to utterance-level authenticity predictions and are evaluated using the three labels defined above.
\begin{itemize}
	\item \textbf{W2V2-AASIST\footnote{https://github.com/TakHemlata/SSL\_Anti-spoofing} \cite{tak2022automatic}:} It combines a pre-trained Wav2Vec 2.0 front-end with a spectro-temporal graph attention network (AASIST) back-end, leveraging heterogeneous attention to capture artifacts across time and frequency domains.
	\item \textbf{W2V2-Conformer\footnote{https://github.com/ErosRos/conformer-based-classifier-for-anti-spoofing} \cite{rosello2023conformer}:} It integrates Wav2Vec 2.0 with a Conformer encoder, where the classification token captures discriminative features, and temporal convolution modules model fine-grained transient anomalies.
	\item \textbf{TCM\footnote{{https://github.com/ductuantruong/tcm\_add}} \cite{truong2024temporal}:} It introduces a Temporal-Channel Modeling (TCM) mechanism that enhances self-attention by jointly modeling temporal and channel dependencies for improved artifact characterization.
	\item \textbf{XLSR-SLS\footnote{https://github.com/QiShanZhang/SLSforASVspoof-2021-DF} \cite{zhang2024audio}:} It leverages a Sensitive Layer Selection (SLS) module to exploit multi-layer representations from the pre-trained XLS-R model, improving robustness through selective contextual modeling.
	\item \textbf{Nes2Net-X\footnote{https://github.com/Liu-Tianchi/Nes2Net} \cite{liu2025nes2net}:} It proposes lightweight, dimensionality reduction (DR)-free architectures that directly process high-dimensional features, reducing overhead while preserving information.
\end{itemize}

\noindent\textbf{SSL-based Methods for Temporal Manipulation Localization.}
Unlike standard detection, these methods predict frame-wise probabilities or exact temporal boundaries.
\begin{itemize}
	\item \textbf{MRM\footnote{https://github.com/nii-yamagishilab/PartialSpoof} \cite{partialspoof}:} It integrates frame- and utterance-level modeling to detect short spoofed segments, enabling precise localization with fine-grained supervision.
	\item \textbf{TDL\footnote{https://github.com/xieyuankun/TDL-ADD} \cite{TDL}:} It proposes a temporal deepfake localization method that separates authentic and synthetic frames in the embedding space via similarity modeling.
	\item \textbf{BAM\footnote{https://github.com/media-sec-lab/BAM} \cite{BAM}:} It introduces a boundary-aware attention mechanism to enhance localization accuracy by explicitly modeling boundary information.
	\item \textbf{CFPRF\footnote{https://github.com/ItzJuny/CFPRF} \cite{CFPRF}:} It presents a coarse-to-fine refinement framework with a temporal localization network to predict precise start and end points of forgery segments.
\end{itemize}

\noindent\textbf{ALLM-based Methods.}
ALLM-based methods formulate audio forensics as an instruction-following text-generation task.
For example, ALLM4ADD\footnote{https://github.com/ucas-hao/qwen\_audio\_for\_add} \cite{ALLM4ADD} casts audio deepfake detection as a question-answering task with ALLMs, enabling robust fake-or-real judgments via supervised fine-tuning, especially in low-data scenarios.

\section{More Implementation Details}
\label{supp:more_imple}

\subsection{Data Preprocessing}
\label{supp:data_prep}
ThinkOmni operates on an omni-modal input space consisting of audio waveforms, textual instructions, and visual spectrograms.
\begin{itemize}
	\item \textbf{Semantic Modality:} The raw audio is processed by the semantic audio encoder retained from Qwen2.5-Omni, which is based on the Whisper-large-v3 architecture and converts speech content into semantic latent representations.
	\item \textbf{Acoustic Modality:} All input audio is resampled to 16~kHz to match the input requirements of the wav2vec~2.0 XLSR acoustic encoder.
	\item \textbf{Visual Modality:} A linear spectrogram is generated using the Short-Time Fourier Transform (STFT) with a window length of 1,024 samples and a hop length of 256 samples. It is converted to the decibel scale and resized to $224\times224$ pixels before being encoded by the vision tower.
\end{itemize}

The waveform and spectrogram are generated from the same audio interval. Resizing the spectrogram changes only its visual resolution and does not redefine the temporal annotations, which remain expressed in the waveform time coordinate system.

\subsection{Model Configuration}
\label{supp:model_conf}
ThinkOmni applies Low-Rank Adaptation (LoRA) to the Thinker backbone so that the language-model parameters can be adapted with a limited number of trainable weights.
\begin{itemize}
	\item \textbf{Target Modules:} LoRA is injected into all linear layers within the Transformer blocks of the LLM backbone, specifically including \texttt{q\_proj}, \texttt{k\_proj}, \texttt{v\_proj}, \texttt{o\_proj}, \texttt{gate\_proj}, \texttt{up\_proj}, and \texttt{down\_proj}.
	\item \textbf{LoRA Hyperparameters:} The LoRA rank is $r=8$, the scaling factor is $\alpha=32$, and the dropout rate is $0.05$ in all compatible ALLM configurations.
	\item \textbf{SAFE Module Dimensions:} In SAFE, semantic features are $D_{sem}=3584$ for the 7B model or $2048$ for the 3B model, and acoustic features are $D_{xlsr}=1024$. Consistent with the notation in Eq.~(\ref{eq:loc}), the cross-attention bottleneck dimension is $D_k=256$.
\end{itemize}

\begin{table}[t]
	\centering
	\caption{Available implementation settings for SFA, AFA, and MFR. Learning Rate$^*$ applies to the Thinker/LLM parameters, whereas Learning Rate$^\Delta$ applies to the non-LLM modules trained in the corresponding stage.}
	\label{tab:train_para}
	\begin{tabular}{cccc}
		\toprule
		\textbf{Parameter} & \textbf{SFA} & \textbf{AFA} & \textbf{MFR} \\
		\midrule
		Batch Size          & 2            & 2            & 2            \\
		Epochs & 1 & 1 & 1 \\
		Warmup Ratio & 0.05 & 0.05 & 0.05 \\
		Learning Rate$^*$   & $1\times10^{-4}$ & $1\times10^{-4}$ & $1\times10^{-4}$ \\
		Learning Rate$^\Delta$ & $1\times10^{-5}$ & $1\times10^{-5}$ & $1\times10^{-5}$ \\
		LoRA Rank & 8 & 8 & 8 \\
		LoRA $\alpha$ & 32 & 32 & 32 \\
		Trainable Parameters & 26.12M & 83.02M & 26.57M \\
		\bottomrule
	\end{tabular}
\end{table}

\begin{table*}[t]
	\centering
	\caption{Comparison of ThinkOmni with SOTA methods for intra- and cross-dataset spoofing detection. P and R denote precision and recall (\%), respectively.}
	\label{tab:det_results}
	\resizebox{\textwidth}{!}{
		\begin{tabular}{ll | cc cc  cc cc cc cc cc cc | cc cc cc}
			\toprule
			\raisebox{-4ex}[0pt][0pt]{Method} & \raisebox{-4ex}[0pt][0pt]{Source} & \multicolumn{16}{c|}{Intra-Dataset} & \multicolumn{6}{c}{Cross-Dataset} \\
			\cmidrule(lr){3-18} \cmidrule(lr){19-24}
			& & \multicolumn{2}{c}{PS} & \multicolumn{2}{c}{HAD} & \multicolumn{2}{c}{LAV-DF} & \multicolumn{2}{c}{SINE} & \multicolumn{2}{c}{LPS} & \multicolumn{2}{c}{ArEnAV} & \multicolumn{2}{c}{AV-1M++} & \multicolumn{2}{c|}{\textbf{Avg.}} & \multicolumn{2}{c}{ADD} & \multicolumn{2}{c}{SF} & \multicolumn{2}{c}{\textbf{Avg.}} \\
			\cmidrule(lr){3-4} \cmidrule(lr){5-6} \cmidrule(lr){7-8}
			\cmidrule(lr){9-10} \cmidrule(lr){11-12} \cmidrule(lr){13-14}
			\cmidrule(lr){15-16} \cmidrule(lr){17-18}
			\cmidrule(lr){19-20} \cmidrule(lr){21-22} \cmidrule(lr){23-24}
			& & P & R & P & R & P & R & P & R & P & R & P & R & P & R & P & R & P & R & P & R & P & R \\
			\midrule
			\multicolumn{24}{l}{\textit{SSL-based Methods}} \\
			W2V2-AASIST & Odyssey'22 & 89.20 & 89.80 & 99.60 & 99.21 & 91.77 & 90.79 & 78.88 & 78.39 & 86.64 & 85.74 & 96.47 & 96.42 & 89.98 & 89.63 & 90.36 & 90.00 & 62.30 & 60.35 & 55.30 & 21.90 & 58.80 & 41.13 \\
			W2V2-Conf. & Interspeech'23 & 92.00 & 91.79 & 99.63 & 99.32 & 95.62 & 95.58 & 82.77 & 82.60 & 89.59 & 88.38 & 96.91 & 96.89 & \underline{93.16} & \underline{93.01} & 92.81 & 92.51 & 66.64 & 66.72 & 62.83 & 26.72 & 64.74 & 46.72 \\
			TCM & Interspeech'24 & 92.75 & 93.08 & 99.69 & \underline{99.44} & 93.00 & 92.28 & \underline{86.93} & \underline{86.72} & \textbf{90.71} & \underline{90.47} & 97.25 & 97.22 & 91.32 & 90.68 & \underline{93.09} & \underline{92.84} & 66.67 & 67.41 & 64.52 & 23.51 & 65.60 & 45.46 \\
			XLSR-SLS & MM'24 & 90.16 & 90.16 & 99.76 & \textbf{99.58} & 95.03 & 95.01 & 82.85 & 82.88 & 88.92 & 87.74 & \underline{96.96} & \underline{96.94} & 92.39 & 92.26 & 92.30 & 92.08 & 67.86 & 68.96 & 57.85 & 20.57 & 62.86 & 44.77 \\
			Nes2Net-X & TIFS'25 & 86.13 & 86.77 & \underline{99.78} & 99.06 & 96.16 & 96.16 & \textbf{88.17} & \textbf{87.86} & 87.80 & 88.39 & \textbf{97.52} & \textbf{97.52} & 92.92 & 92.87 & 92.64 & 92.66 & 70.45 & 71.02 & 55.33 & 13.68 & 62.89 & 42.35 \\
			\midrule
			\multicolumn{24}{l}{\textit{ALLM-based Methods}} \\
			ALLM4ADD & MM'25 & \textbf{96.48} & \textbf{96.48} & \textbf{99.92} & 98.39 & 96.18 & 95.89 & 64.59 & 62.73 & 90.35 & 90.07 & 94.92 & 94.88 & 90.79 & 90.04 & 90.46 & 89.78 & 75.42 & 72.61 & \underline{98.38} & 51.96 & 86.90 & 62.29 \\
			Qwen2-Audio & - & 87.23 & 84.21 & 99.65 & 90.04 & \underline{96.95} & \underline{96.92} & 63.27 & 59.97 & 73.60 & 65.75 & 92.23 & 92.14 & 91.39 & 91.36 & 86.33 & 82.91 & 70.76 & 69.18 & 88.31 & \textbf{83.15} & 79.54 & \underline{76.17} \\
			Qwen2.5-Omni-3B  & - & 88.25 & 87.31 & 99.75 & 90.13 & 96.76 & 96.73 & 75.24 & 72.99 & 84.79 & 82.51 & 92.10 & 92.08 & 90.05 & 89.93 & 89.56 & 87.38 & \underline{84.26} & \underline{75.33} & 93.57 & 47.05 & 88.92 & 61.19 \\
			Qwen2.5-Omni-7B  & - & 82.79 & 81.15 & 99.72 & 93.58 & 93.69 & 93.32 & 70.02 & 63.15 & 71.65 & 64.78 & 90.52 & 90.44 & 85.84 & 84.68 & 84.89 & 81.59 & 83.70 & 75.17 & 95.83 & 62.05 & \underline{89.77} & 68.61 \\
			\midrule
			\textbf{ThinkOmni} & \textbf{Ours} & \underline{94.18} & \underline{93.87} & \underline{99.78} & 98.23 & \textbf{99.46} & \textbf{99.46} & 81.96 & 81.96 & \underline{90.65} & \textbf{90.64} & 96.63 & 96.51 & \textbf{95.25} & \textbf{95.24} & \textbf{93.99} & \textbf{93.70} & \textbf{84.83} & \textbf{78.87} & \textbf{98.53} & \underline{82.61} & \textbf{91.68} & \textbf{80.74} \\
			\bottomrule
		\end{tabular}
	}
\end{table*}

\begin{table*}[t]
	\centering
	\caption{Comparison of ThinkOmni with SOTA methods for temporal manipulation localization across both intra- and cross-dataset settings under different IoU thresholds.}
	\label{tab:loc_results}
	\resizebox{\textwidth}{!}{
		\begin{tabular}{c l l|cccccccc|ccc||cccccccc|ccc}
			\toprule
			\raisebox{-2ex}[0pt][0pt]{AP@} & \raisebox{-2ex}[0pt][0pt]{Method} & \raisebox{-2ex}[0pt][0pt]{Source} 
			& \multicolumn{8}{c|}{Intra-Dataset} & \multicolumn{3}{c||}{Cross-Dataset}
			& \multicolumn{8}{c|}{Intra-Dataset} & \multicolumn{3}{c}{Cross-Dataset} \\
			\cmidrule(lr){4-11} \cmidrule(lr){12-14}
			\cmidrule(lr){15-22} \cmidrule(lr){23-25}
			&  &  & PS & HAD & LAV-DF & SINE & LPS & ArEnAV & AV-1M++ & Avg. & ADD & SF & Avg.
			& PS & HAD & LAV-DF & SINE & LPS & ArEnAV & AV-1M++ & Avg. & ADD & SF & Avg. \\
			\midrule
			
			\raisebox{-16ex}[0pt][0pt]{\shortstack{0.5 \\ / \\ 0.75}}
			& \multicolumn{24}{l}{\textit{SSL-based Method}} \\
			& MRM & TASLP'23 
			& 39.22 & 94.83 & 89.45 & 31.73 & 44.29 & \underline{97.32} & 84.92 & 68.82 & 1.64 & 0.17 & 0.91
			& 30.92 & 87.66 & 86.24 & 23.58 & 31.69 & \underline{93.86} & 77.16 & 61.59 & 0.35 & 0.02 & 0.19 \\
			& TDL & ICASSP'24 
			& \textbf{91.03} & 97.22 & \underline{97.87} & 63.42 & \textbf{91.26} & 96.35 & 90.30 & \underline{89.64} & 62.91 & 6.31 & 34.61
			& \textbf{80.19} & 84.05 & 90.09 & 57.18 & 77.25 & 88.61 & 76.32 & 79.10 & 58.52 & 1.99 & 30.26 \\
			& BAM & Interspeech'24 
			& 58.18 & \underline{99.30} & 94.68 & 67.78 & 86.38 & 88.07 & \textbf{96.78} & 84.45 & 0.29 & 5.57 & 2.93
			& 50.35 & \textbf{98.46} & 90.17 & 60.76 & \textbf{78.51} & 85.66 & \textbf{93.70} & 79.66 & 0.06 & 4.09 & 2.08 \\
			& CFPRF & MM'24 
			& 65.47 & \textbf{99.81} & 92.68 & 77.06 & 79.03 & 96.31 & 81.86 & 84.60 & 4.03 & 6.52 & 5.28
			& 55.97 & \underline{95.25} & 88.53 & 65.54 & 67.38 & 89.64 & 60.33 & 74.66 & 0.56 & 3.03 & 1.80 \\
			\cmidrule(lr){2-25}
			
			& \multicolumn{24}{l}{\textit{ALLM-based Method}} \\
			& Qwen-Audio & - 
			& 70.73 & 58.82 & 77.53 & 43.12 & 69.41 & 78.72 & 55.61 & 64.85 & 37.58 & 2.10 & 19.84
			& 67.03 & 30.42 & 62.85 & 40.53 & 58.39 & 59.43 & 46.38 & 52.15 & 31.26 & 1.51 & 16.39 \\
			& Qwen2-Audio & - 
			& 73.97 & 91.30 & 96.36 & 78.91 & 76.75 & 92.27 & 80.87 & 84.35 & 76.97 & \underline{54.52} & \underline{65.75}
			& 67.20 & 88.08 & 91.33 & 70.89 & 61.70 & 75.04 & 68.29 & 74.65 & 67.68 & \underline{51.10} & \underline{59.39} \\
			& Qwen2.5-Omni-3B & - 
			& 81.22 & 81.66 & 93.83 & 83.92 & 82.09 & 93.68 & 83.91 & 85.76 & \underline{81.73} & 15.72 & 48.73
			& 77.09 & 78.86 & 91.77 & 81.60 & 72.16 & 89.43 & 78.38 & 81.33 & 73.24 & 12.30 & 42.77 \\
			& Qwen2.5-Omni-7B & - 
			& \underline{81.38} & 86.56 & 95.80 & \textbf{95.63} & 76.24 & 93.60 & 86.05 & 87.89 & 75.27 & 47.91 & 61.59
			& \underline{77.14} & 84.99 & \underline{94.17} & \textbf{94.54} & 65.75 & 91.29 & 80.21 & \underline{84.01} & \underline{67.77} & 43.92 & 55.85 \\
			& ThinkOmni & Ours 
			& 78.94 & 94.71 & \textbf{98.23} & \underline{88.40} & \underline{86.69} & \textbf{98.07} & \underline{94.33} & \textbf{91.34} & \textbf{83.02} & \textbf{80.68} & \textbf{81.85}
			& 73.68 & 93.97 & \textbf{96.60} & \underline{85.85} & \underline{78.14} & \textbf{97.05} & \underline{92.15} & \textbf{88.21} & \textbf{73.05} & \textbf{76.40} & \textbf{74.73} \\
			
			\midrule
			
			\raisebox{-16ex}[0pt][0pt]{\shortstack{0.9 \\ / \\ 0.95}}
			& \multicolumn{24}{l}{\textit{SSL-based Method}} \\
			& MRM & TASLP'23 
			& 26.50 & 82.48 & 66.31 & 16.21 & 23.33 & 83.67 & 74.48 & 53.28 & 0.05 & 0.01 & 0.03
			& 25.68 & 81.99 & 65.81 & 16.14 & 21.09 & 83.01 & 74.46 & 52.60 & 0.05 & 0.01 & 0.03 \\
			& TDL & ICASSP'24 
			& 72.67 & 57.99 & 72.85 & 55.66 & 68.18 & 74.49 & 69.71 & 67.36 & 57.74 & 1.32 & 29.53
			& 71.32 & 56.16 & 71.65 & 55.66 & 66.85 & 73.91 & 69.47 & 66.43 & 57.72 & 1.29 & 29.51 \\
			& BAM & Interspeech'24 
			& 45.86 & \textbf{98.02} & 80.34 & 57.85 & \underline{72.32} & 84.70 & \underline{89.49} & 75.51 & 0.05 & 3.56 & 1.81
			& 43.17 & \textbf{97.91} & 80.28 & 57.73 & \underline{70.80} & \underline{84.70} & \textbf{89.36} & 74.85 & 0.05 & 3.53 & 1.79 \\
			& CFPRF & MM'24 
			& 49.79 & 83.08 & 59.75 & 57.31 & 56.25 & 66.77 & 50.75 & 60.53 & 0.08 & 1.01 & 0.55
			& 45.96 & 78.67 & 56.89 & 51.79 & 42.35 & 62.48 & 49.18 & 55.33 & 0.08 & 0.87 & 0.48 \\
			\cmidrule(lr){2-25}
			
			& \multicolumn{24}{l}{\textit{ALLM-based Method}} \\
			& Qwen-Audio & - 
			& 64.34 & 12.09 & 51.38 & 39.95 & 54.53 & 48.45 & 43.60 & 44.91 & 29.77 & 1.36 & 15.57
			& 62.21 & 7.18 & 49.12 & 39.89 & 52.32 & 46.70 & 43.26 & 42.95 & 29.67 & 1.34 & 15.51 \\
			& Qwen2-Audio & - 
			& 63.72 & 80.00 & 79.16 & 56.96 & 56.47 & 58.07 & 55.99 & 64.34 & 62.16 & \underline{47.59} & \underline{54.88}
			& 60.36 & 60.54 & 68.75 & 51.76 & 53.60 & 53.70 & 52.23 & 57.28 & 56.63 & \underline{45.17} & \underline{50.90} \\
			& Qwen2.5-Omni-3B & - 
			& \underline{74.62} & 77.67 & 88.80 & 76.17 & 67.69 & 81.61 & 67.41 & 76.28 & \underline{67.68} & 9.45 & 38.57
			& \underline{73.61} & 73.22 & 81.69 & 72.28 & 66.26 & 71.55 & 59.88 & 71.21 & \underline{66.11} & 5.81 & 35.96 \\
			& Qwen2.5-Omni-7B & - 
			& \textbf{75.51} & 84.23 & \underline{91.98} & \textbf{92.32} & 61.44 & \underline{87.63} & 69.33 & \underline{80.35} & 65.96 & 39.05 & 52.51
			& \textbf{75.12} & 80.98 & \underline{85.03} & \textbf{90.88} & 59.99 & 80.09 & 60.75 & \underline{76.12} & 65.19 & 31.82 & 48.51 \\
			& ThinkOmni & Ours 
			& 70.97 & \underline{93.75} & \textbf{95.35} & \underline{79.36} & \textbf{73.76} & \textbf{95.35} & \textbf{89.64} & \textbf{85.45} & \textbf{69.25} & \textbf{71.12} & \textbf{70.19}
			& 69.60 & \underline{93.39} & \textbf{90.62} & \underline{73.27} & \textbf{71.59} & \textbf{92.97} & \underline{85.77} & \textbf{82.46} & \textbf{68.41} & \textbf{58.64} & \textbf{63.53} \\
			
			\bottomrule
		\end{tabular}
	}
\end{table*}

\subsection{Detailed Training Strategy}
\label{supp:train_stra}
Forensic-Aware Modality-Incremental Learning (FMIL) is implemented as three sequential stages using \texttt{ms-swift}\footnote{https://github.com/modelscope/ms-swift} to configure stage-specific trainable modules and learning rates.

\begin{enumerate}
	\item \textbf{Stage 1: Semantic Forensic Adaptation (SFA).} 
	\begin{itemize}
		\item \textit{Objective:} Adapt the semantic pathway and Thinker to FACoT reasoning supervision before introducing the acoustic and visual pathways.
		\item \textit{Trainable:} LoRA modules of the LLM backbone, and the semantic encoder's projector.
		\item \textit{Inactive/Frozen:} The vision pathway is not optimized in SFA, and the acoustic encoder and SAFE are introduced only in AFA.
	\end{itemize}
	\item \textbf{Stage 2: Acoustic Forensic Augmentation (AFA).}
	\begin{itemize}
		\item \textit{Objective:} Incorporate fine-grained acoustic evidence into the semantic reasoning pathway.
		\item \textit{Trainable:} Acoustic Encoder (last 24 layers via learnable weighted sum), the newly initialized SAFE module (fully tuned), and LoRA modules of the LLM backbone.
		\item \textit{Frozen/Inactive:} The semantic encoder is frozen so that AFA retains the Stage-1 semantic feature extractor, and the vision pathway remains inactive until MFR.
	\end{itemize}
	\item \textbf{Stage 3: Multi-modal Forensic Refinement (MFR).}
	\begin{itemize}
		\item \textit{Objective:} Incorporate spectrogram-based visual evidence for cross-modal verification and temporal boundary prediction.
		\item \textit{Trainable:} Vision encoder, vision-to-LLM aligner, and LoRA modules of the LLM backbone.
		\item \textit{Frozen:} The semantic encoder, acoustic encoder, and SAFE module are frozen. MFR therefore updates the visual pathway and Thinker LoRA modules while retaining the previously learned semantic--acoustic feature extractors.
	\end{itemize}
\end{enumerate}

As shown in Table~\ref{tab:train_para}, each FMIL stage is trained for one epoch and initialized from the preceding checkpoint. Previously trained encoders are frozen to preserve learned features, while the Thinker LoRA modules remain trainable for cross-modal adaptation. This strategy mitigates modality interference.

\subsection{Inference Configuration}
\label{supp:inf_conf}
During inference, ThinkOmni generates outputs autoregressively using greedy decoding for deterministic prediction, with a maximum generation length of 2,048 tokens. For SFA-stage models, inference is performed with vLLM using a maximum context length of 4,096 tokens to accommodate longer semantic reasoning sequences.

The response is parsed into the three fields defined in the main paper: the forensic rationale, the utterance-level detection result, and the localization result. Detection uses labels 0, 1, and 2 for fully real, fully fake, and partially fake audio, respectively. The timestamp-token sequence is converted by $g(\cdot)$ into the predicted interval set $\hat{\mathcal{B}}=\{[\hat{s}_j,\hat{e}_j]\}_{j=1}^{\hat K}$. The exact textual delimiter and ordering rule for multiple intervals must match the training targets and evaluation parser.

\section{More Experimental Results}
\label{supp:more_results}
Detection accuracy is computed over the three authenticity classes, and F1, precision, and recall are support-weighted across these classes. Under this definition, weighted recall is numerically equal to accuracy, while weighted precision and weighted F1 remain distinct. Intra-dataset averages are computed over the seven reported evaluation groups, whereas cross-dataset averages are computed over ADD and Speech-Forensics. Localization mAP is averaged over temporal IoU thresholds from 0.5 to 0.95 in increments of 0.05, following the protocol stated in the main paper.

\subsection{Detection Results}
\label{supp:det_results}
To better illustrate model performance, Table~\ref{tab:det_results} presents Precision (P) and Recall (R), highlighting the trade-offs between avoiding false alarms and reducing missed detections.

\noindent
\textbf{Intra-dataset Performance.}
SSL-based methods are comparatively stable on the intra-dataset groups, whereas several ALLM baselines vary substantially across datasets; for example, ALLM4ADD obtains 64.59\% weighted precision and 62.73\% weighted recall on SINE. ThinkOmni achieves the highest intra-dataset average weighted precision (93.99\%) and an average weighted recall of 93.70\%. These results support the effectiveness of progressive multi-modal learning.

\noindent
\textbf{Cross-dataset Performance.}
Under cross-dataset evaluation, the baselines exhibit substantial degradation under distribution shift. On SF, the weighted recall of the SSL methods decreases to 13\%--27\% (e.g., 13.68\% for Nes2Net-X), indicating sensitivity to training-distribution artifacts. Existing ALLMs are generally more robust but can remain imbalanced; for example, Qwen2.5-Omni-3B obtains 93.57\% weighted precision and 47.05\% weighted recall on SF. ThinkOmni achieves 91.68\% average weighted precision and 80.74\% average weighted recall across the two cross-dataset test sets.
On SF, ThinkOmni obtains 98.53\% weighted precision and 82.61\% weighted recall. 
Together with the reasoning ablations, these results indicate that explicit CoT supervision and progressive multi-modal adaptation contribute to more stable decisions under the evaluated acoustic variations and manipulation types.

\subsection{Localization Results}
\label{supp:loc_results}
Table~\ref{tab:loc_results} presents a fine-grained evaluation of Average Precision (AP) at multiple IoU thresholds (0.5, 0.75, 0.9, and 0.95), assessing the models’ ability to predict precise temporal boundaries rather than approximate localizations.

\noindent
\textbf{Intra-dataset Performance.}
Performance declines as the IoU threshold increases. For example, TDL drops from 89.64\% AP@0.5 to 66.43\% AP@0.95 in the intra-dataset setting. ThinkOmni achieves the highest average AP across all thresholds, including 82.46\% at AP@0.95, demonstrating more precise temporal boundary alignment on the source-domain test sets.

\noindent
\textbf{Cross-dataset Performance.}
Cross-dataset temporal localization further reveals sensitivity to unseen conditions. Several SSL methods approach 0\% AP on SF at strict IoU thresholds, while the ALLM baselines retain higher but still limited boundary precision; for example, Qwen2-Audio achieves 50.90\% cross-dataset average AP@0.95.
ThinkOmni achieves 81.85\% cross-dataset average AP@0.5 and 63.53\% AP@0.95, including 58.64\% AP@0.95 on SF. These are the highest values among the methods reported in the table. The component ablations, rather than this comparison alone, provide evidence about the contributions of CoT supervision and adaptive localization loss.

\begin{table}[t]
	\centering
	\caption{Ablation of token-weighting factors under cross-dataset evaluation. $(\omega_{think},\alpha_{det},\omega_{loc})=(0,0,0)$ denotes standard cross-entropy without role-specific token weighting.}
	\label{tab:token_ablation}
	\begin{tabular}{c c c|c c c c}
		\toprule
		$\omega_{think}$ & $\alpha_{det}$ & $\omega_{loc}$ & mACC & mF1 & mAP & Avg. \\
		\midrule
		0.0 & 0.0 & 0.0 & \underline{68.61} & \underline{75.23} & 55.91 & 66.58 \\
		0.1 & 0.1 & 0.1 & 66.54 & 70.94 & \textbf{71.36} & 69.61 \\
		0.1 & 0.1 & 0.8 & 66.18 & 71.65 & 66.24 & 68.02 \\
		0.1 & 0.4 & 0.5 & 67.64 & 72.56 & 67.88 & 69.36 \\
		0.2 & 0.2 & 0.6 & \textbf{69.79} & \textbf{75.64} & 70.82 & \textbf{72.08} \\
		0.3 & 0.2 & 0.5 & 67.38 & 72.24 & \underline{71.27} & \underline{70.30} \\
		0.4 & 0.2 & 0.4 & 67.15 & 73.15 & 63.25 & 67.85 \\
		\bottomrule
	\end{tabular}
\end{table}

\subsection{Effect of Token Weighting}
\label{supp:token_results}
We examine the association between the role-specific token weights $(\omega_{think},\alpha_{det},\omega_{loc})$ and cross-dataset performance in Table~\ref{tab:token_ablation}. With role-specific weighting disabled, the model obtains 68.61\% mACC, 75.23\% mF1, and 55.91\% mAP. The lower mAP is consistent with the motivation that numerous reasoning tokens can imbalance a sequence loss, but this table alone does not prove that reasoning-token length is the sole cause of the localization gap.
Role-specific weighting improves the best reported overall average from 66.58\% to 72.08\%. The configuration $(\omega_{think},\alpha_{det},\omega_{loc})=(0.2,0.2,0.6)$ provides the highest mACC, mF1, and overall average, whereas $(0.1,0.1,0.1)$ provides the highest mAP. Thus, the ablation shows a trade-off rather than establishing that one weight is independently responsible for all gains. 

\subsection{Computational Efficiency}
\label{supp:computational_efficiency}

Despite incorporating an additional acoustic encoder and SAFE module, ThinkOmni introduces only marginal computational overhead. As shown in Table~\ref{tab:inference_cost}, compared with Qwen2.5-Omni-7B, ThinkOmni increases the parameter count by 4.6\% and computational cost by only 0.5\%, while requiring merely 0.75~GiB additional peak GPU memory and 0.02~s additional inference latency. These modest increases demonstrate a favorable efficiency--performance trade-off, as ThinkOmni achieves substantial cross-dataset gains in both spoofing detection and temporal localization, as reported in Tables~\ref{tab:det_results} and~\ref{tab:loc_results}.

\begin{table}[t]
	\centering
	\caption{Computational costs of ThinkOmni and Qwen2.5-Omni baselines. Peak GPU memory is measured with a batch size of 4 under identical hardware, precision, input, and decoding settings.}
	\label{tab:inference_cost}
	\resizebox{\linewidth}{!}{
		\begin{tabular}{lcccc}
			\toprule
			Method
			& Params. (B)
			& FLOPs (T)
			& Memory (GiB)
			& Latency (s) \\
			
			\midrule
			Qwen2.5-Omni-3B \cite{Qwen2.5-Omni}
			& 5.54
			& 110.51
			& 12.04
			& 1.42 \\
			
			Qwen2.5-Omni-7B \cite{Qwen2.5-Omni}
			& 8.93
			& 220.05
			& 22.62
			& 2.82 \\
			
			\textbf{ThinkOmni} & 9.34 & 221.19 & 23.37 & 2.84 \\
			\bottomrule
	\end{tabular}}
\end{table}

\begin{table}[t]
	\centering
	\caption{Stage-wise comparison of models trained and evaluated with and without CoT under cross-dataset evaluation. All values are reported in percent.}
	\label{tab:cot_stage_ablation}
	\begin{tabular}{lcccccc}
		\toprule
		\raisebox{-2.5ex}[0pt][0pt]{Metric}
		& \multicolumn{3}{c}{With CoT}
		& \multicolumn{3}{c}{Without CoT} \\
		\cmidrule(lr){2-4}
		\cmidrule(lr){5-7}
		& SFA & AFA & MFR
		& SFA & AFA & MFR \\
		\midrule
		mACC $\uparrow$
		& 69.79 & \textbf{74.55} & \textbf{80.74}
		& \textbf{71.06} & 73.20 & 68.12 \\
		
		mF1 $\uparrow$
		& 75.64 & \textbf{80.34} & \textbf{85.15}
		& \textbf{76.76} & 80.09 & 74.64 \\
		
		mAP $\uparrow$
		& \textbf{70.82} & \textbf{72.26} & \textbf{74.67}
		& 66.34 & 66.77 & 63.57 \\
		\bottomrule
	\end{tabular}
\end{table}

\subsection{Stage-wise Effect of CoT}
\label{supp:cot_stage_ablation}
We compare models trained and evaluated with and without CoT across the SFA, AFA, and MFR stages under identical data and architectures. The two settings differ only in whether structured rationales are used during training and inference.

As shown in Table~\ref{tab:cot_stage_ablation}, CoT yields a detection--localization trade-off at SFA, reducing mACC and mF1 by 1.27\% and 1.12\% while improving mAP by 4.48\%. At AFA, it improves mACC, mF1, and mAP by 1.35\%, 0.25\%, and 5.49\%, respectively. The gains further increase at MFR to 12.62\%, 10.51\%, and 11.10\%.

These results indicate that CoT becomes increasingly effective as acoustic and spectral-visual cues are incorporated, facilitating multi-modal forensic reasoning and temporal localization.

\subsection{Effect of SAFE Fusion}
\label{supp:safe_ablation}
We further evaluate the contribution of SAFE by replacing it with direct feature concatenation under the same SFA+AFA training setting and the same XLSR-300M acoustic encoder. As shown in Table~\ref{tab:safe_ablation}, SAFE consistently outperforms naive concatenation across all detection and localization metrics.

On the intra-dataset test sets, SAFE improves mACC, mF1, and mAP by 26.41\%, 24.65\%, and 6.16\%, respectively. Under cross-dataset evaluation, the corresponding improvements are 18.22\%, 12.39\%, and 2.05\%. The particularly large gains in mACC and mF1 show that direct concatenation is insufficient for reconciling heterogeneous semantic and acoustic representations. Meanwhile, the consistent mAP improvements indicate that SAFE also preserves fine-grained evidence useful for temporal boundary prediction. These controlled results demonstrate the effectiveness of SAFE for semantic--acoustic forensic fusion.

\begin{table}[t]
	\centering
	\caption{Ablation of SAFE under the matched SFA+AFA setting with XLSR-300M.}
	\label{tab:safe_ablation}
	\resizebox{\linewidth}{!}{
		\begin{tabular}{lcccccc}
			\toprule
			\raisebox{-2.5ex}[0pt][0pt]{Method}
			& \multicolumn{3}{c}{Intra-dataset}
			& \multicolumn{3}{c}{Cross-dataset} \\
			\cmidrule(lr){2-4}
			\cmidrule(lr){5-7}
			& mACC $\uparrow$
			& mF1 $\uparrow$
			& mAP $\uparrow$
			& mACC $\uparrow$
			& mF1 $\uparrow$
			& mAP $\uparrow$ \\
			\midrule
			Concatenation
			& 67.47
			& 69.31
			& 81.63
			& 56.33
			& 67.95
			& 70.21 \\
			
			\textbf{SAFE}
			& \textbf{93.88}\,\textcolor{red}{\scriptsize(+26.41)}
			& \textbf{93.96}\,\textcolor{red}{\scriptsize(+24.65)}
			& \textbf{87.79}\,\textcolor{red}{\scriptsize(+6.16)}
			& \textbf{74.55}\,\textcolor{red}{\scriptsize(+18.22)}
			& \textbf{80.34}\,\textcolor{red}{\scriptsize(+12.39)}
			& \textbf{72.26}\,\textcolor{red}{\scriptsize(+2.05)} \\
			\bottomrule
	\end{tabular}}
\end{table}
\section{Case Study}
\label{supp:case_study}

\subsection{Successful Case Studies}
\label{supp:success_cases}
\begin{figure}[t]
	\centering
	\includegraphics[width=\linewidth]{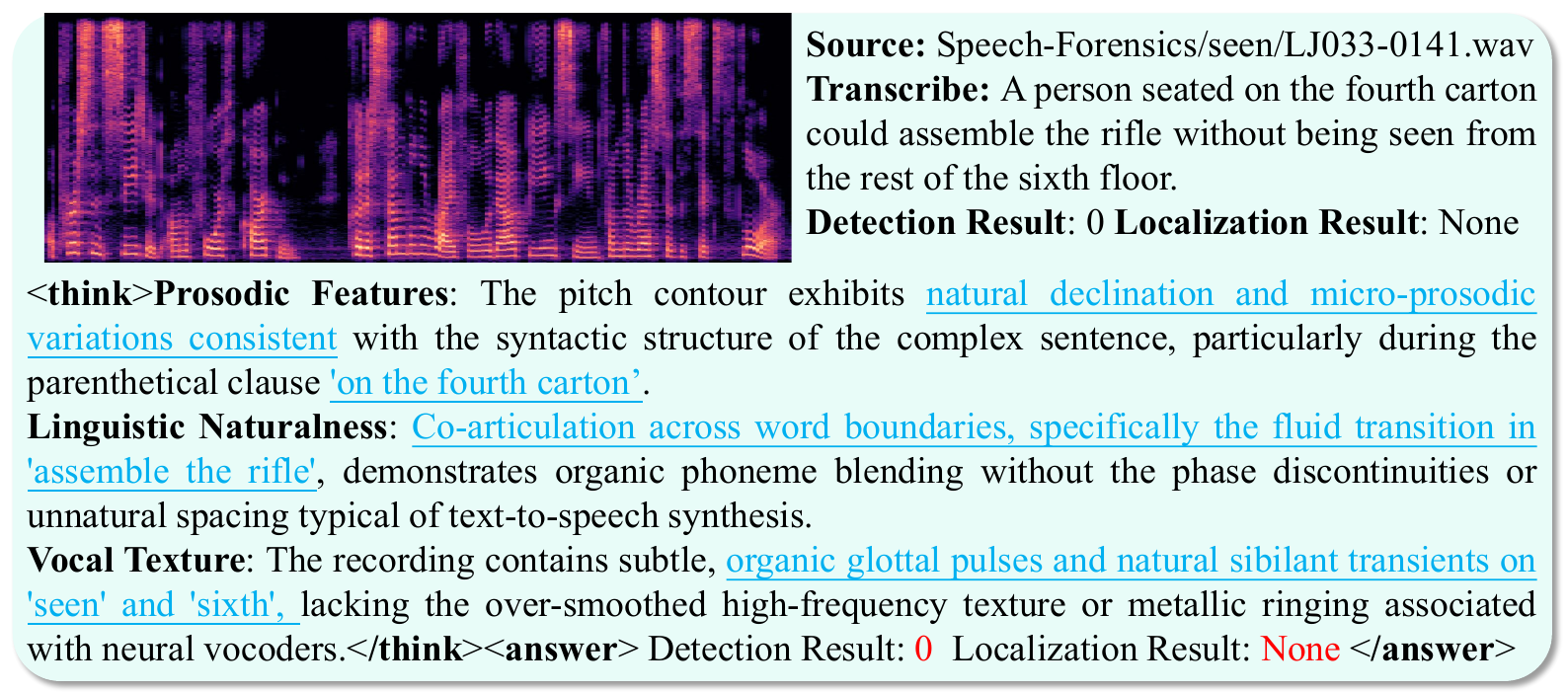}
	\caption{Successful case analysis of a fully real sample.}
	\label{fig:success_fully_real}
\end{figure}

\begin{figure}[t]
	\centering
	\includegraphics[width=\linewidth]{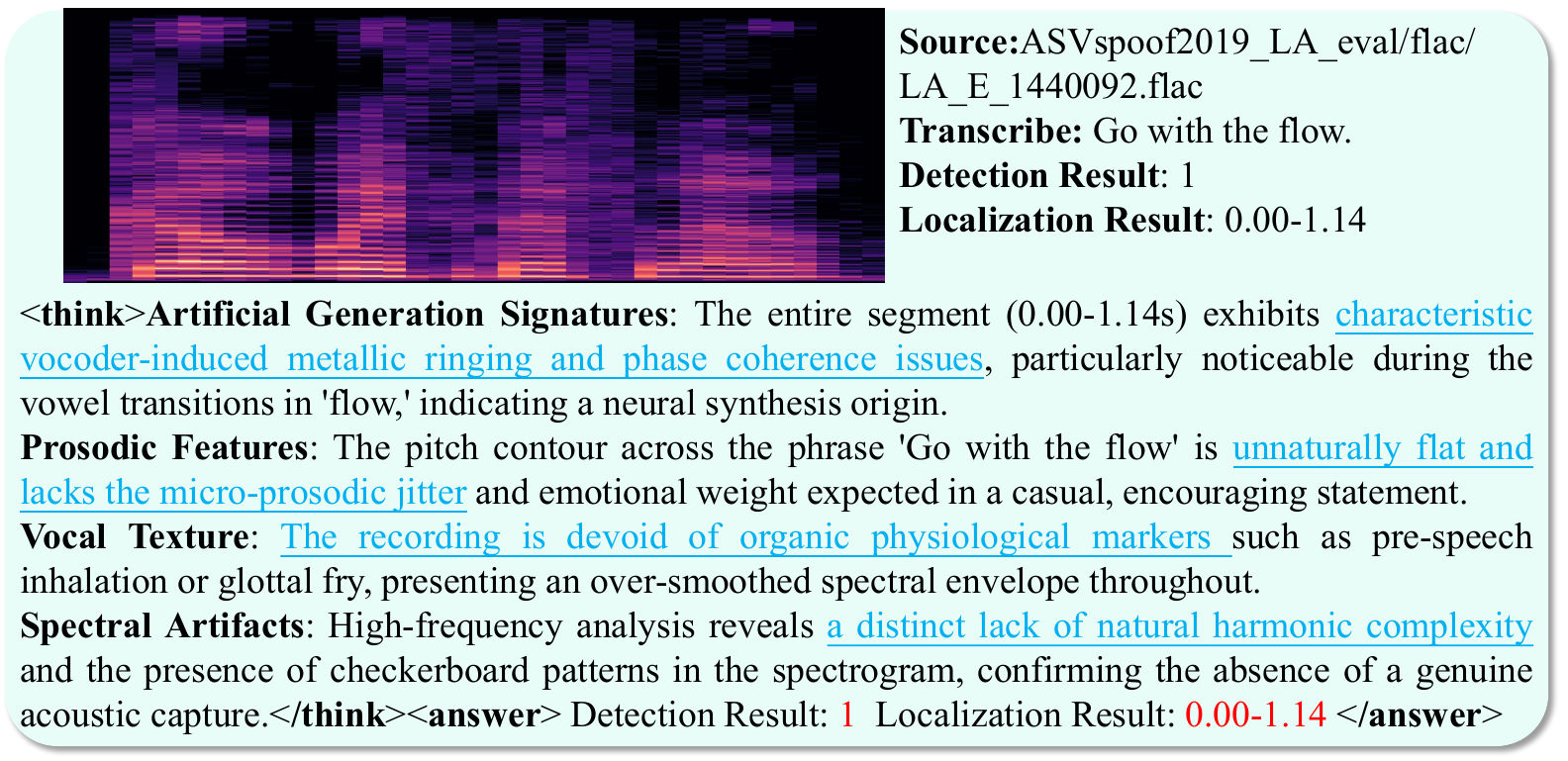}
	\caption{Successful case analysis of a fully fake sample.}
	\label{fig:success_fully_fake}
\end{figure}

\begin{figure}[t]
	\centering
	\includegraphics[width=\linewidth]{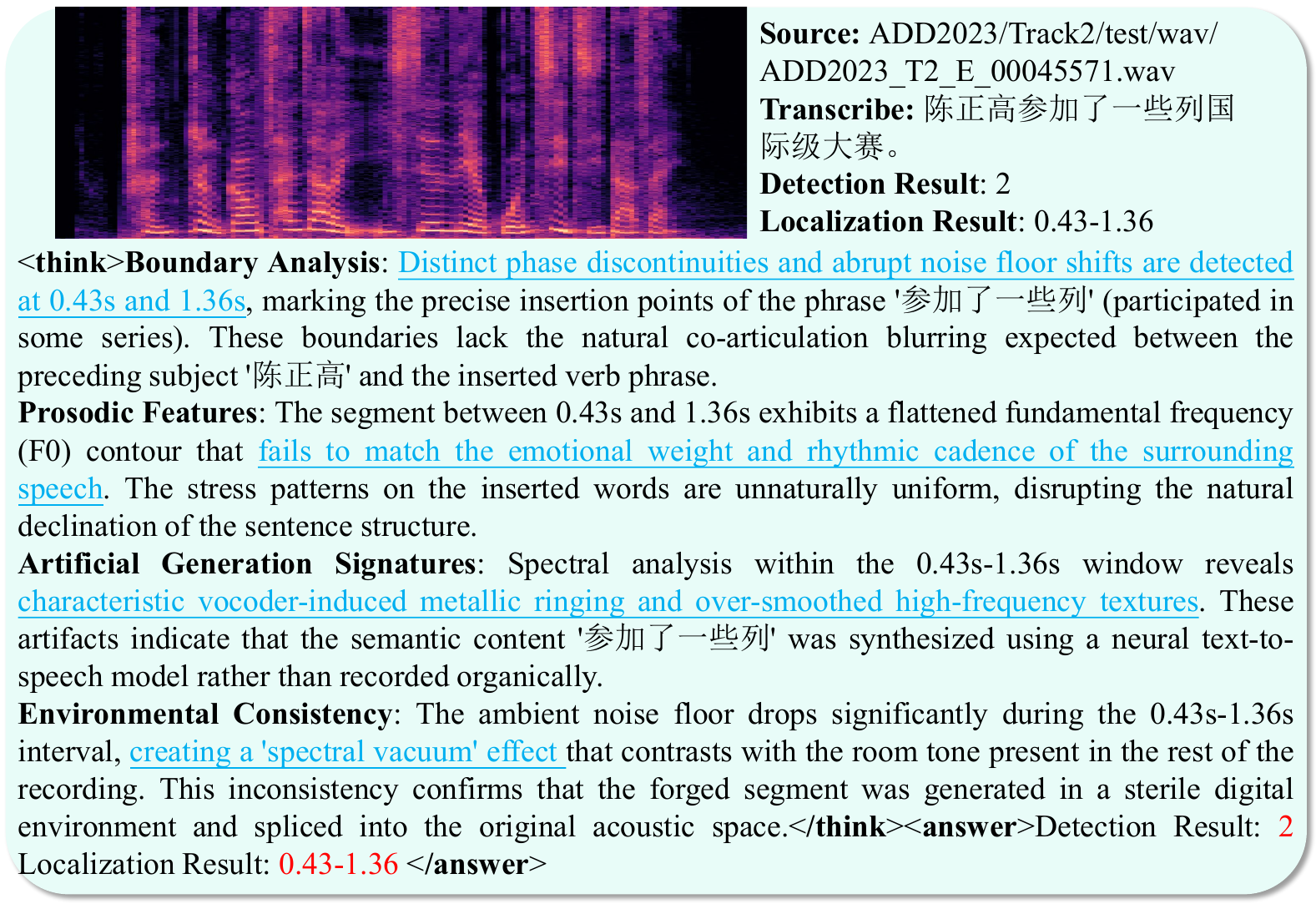}
	\caption{Successful case analysis of a partially fake sample.}
	\label{fig:success_part_fake}
\end{figure}

To illustrate the model outputs, we qualitatively analyze three representative audio samples.
\begin{itemize}
	\item \textbf{Fully Real (Figure~\ref{fig:success_fully_real}):} The model correctly classifies the sample as fully real and cites ``micro-prosodic variations'' and ``organic glottal pulses'' as evidence consistent with natural speech.
	\item \textbf{Fully Fake (Figure~\ref{fig:success_fully_fake}):} The model correctly classifies the entire clip as fully fake and attributes the decision to cues described as ``vocoder-induced metallic ringing'' and ``unnaturally flat'' prosody.
	\item \textbf{Partially Fake (Figure~\ref{fig:success_part_fake}):} The model correctly localizes the annotated manipulated region at 0.43--1.36~s and associates it with reported phase discontinuities, noise-floor shifts, and rhythmic or emotional mismatches. These descriptions summarize the generated rationale and do not independently verify that each cited cue caused the prediction.
\end{itemize}

These examples illustrate how ThinkOmni organizes acoustic, prosodic, environmental, and semantic cues into an inspectable rationale while producing detection and localization outputs. They are qualitative examples and do not establish expert-level reliability on their own.

\subsection{Failure Case Studies}
\label{supp:failed_cases}
\begin{figure}[t]
	\centering
	\includegraphics[width=\linewidth]{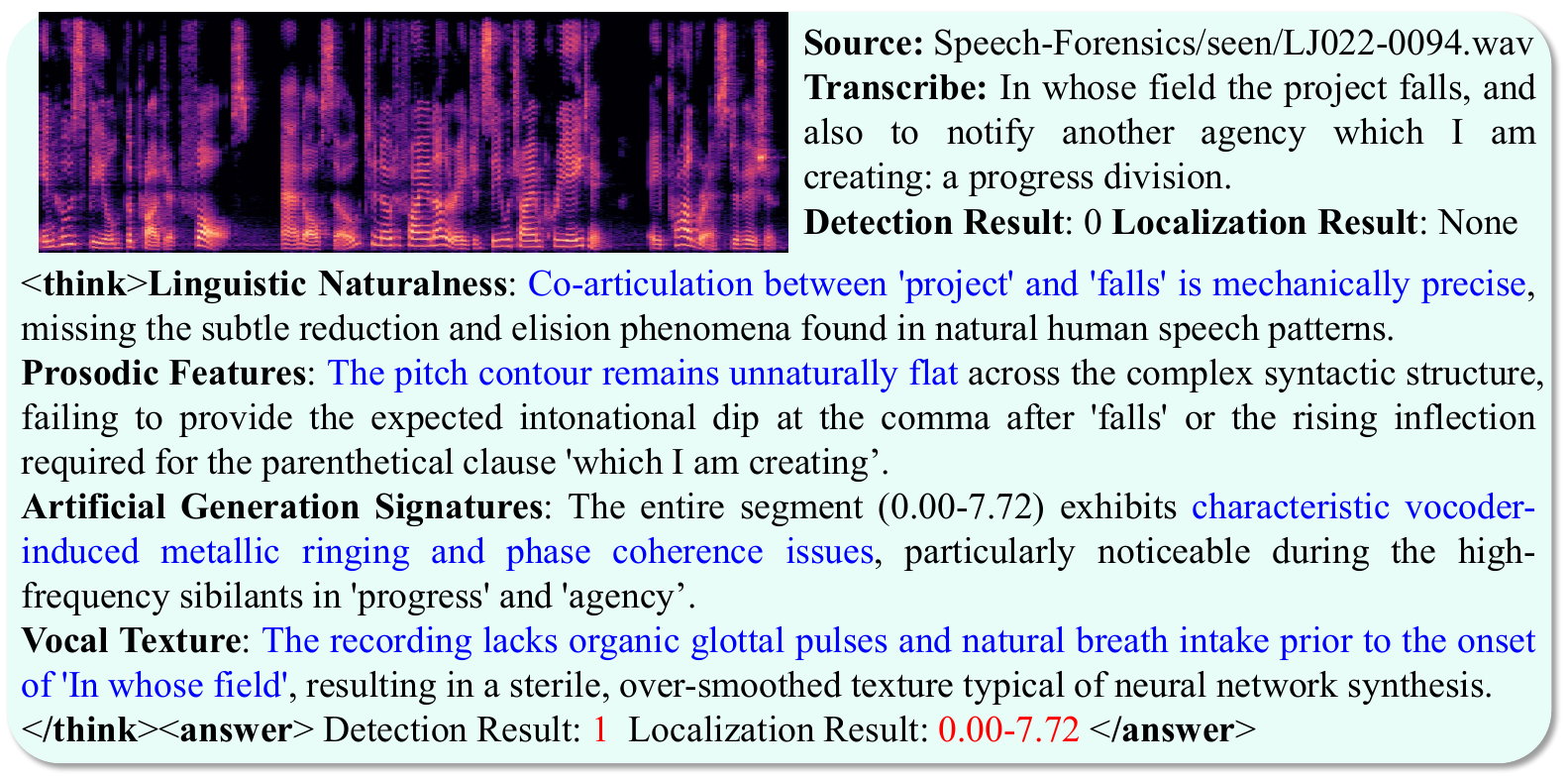}
	\caption{Failure case analysis of a fully real sample.}
	\label{fig:fail_fully_real}
\end{figure}

\begin{figure}[t]
	\centering
	\includegraphics[width=\linewidth]{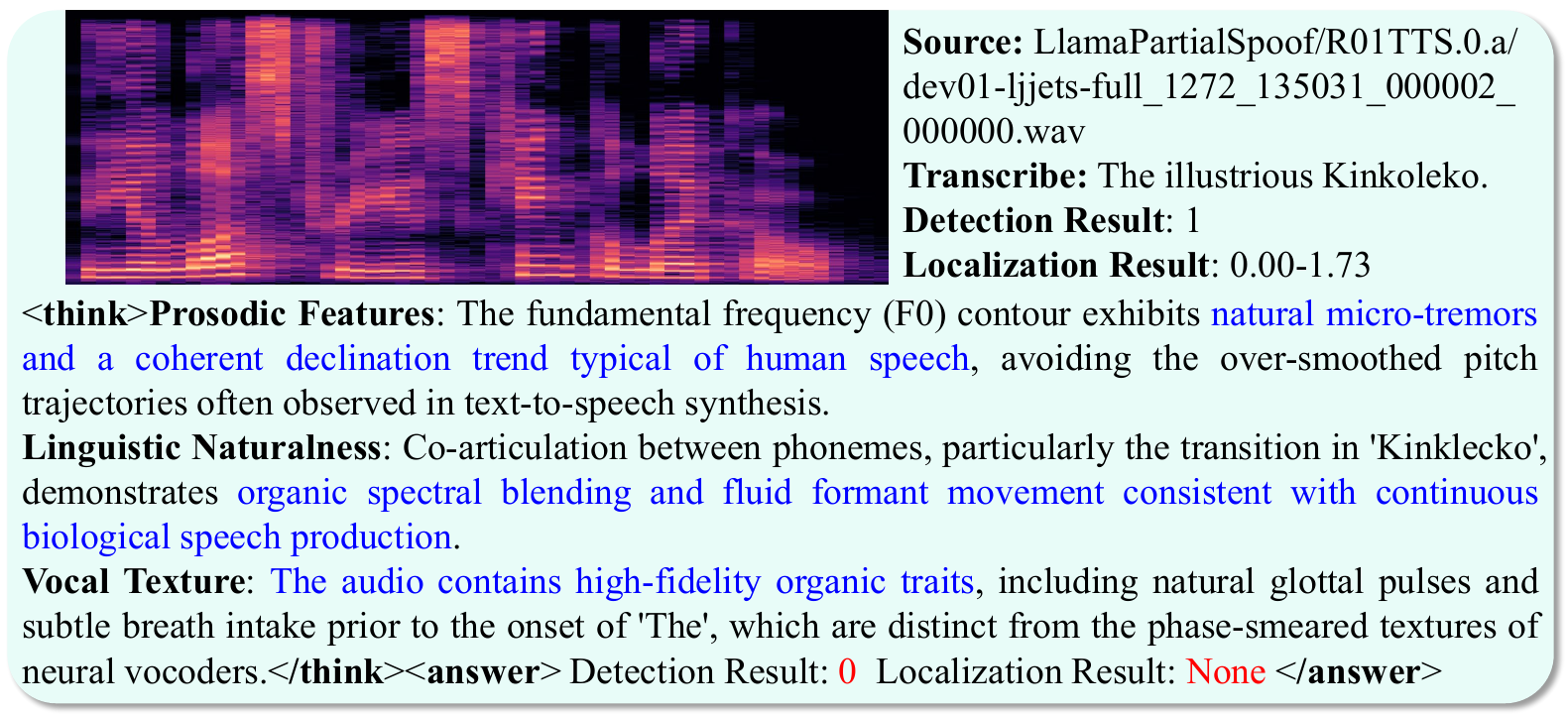}
	\caption{Failure case analysis of a fully fake sample.}
	\label{fig:fail_fully_fake}
\end{figure}

\begin{figure}[t]
	\centering
	\includegraphics[width=\linewidth]{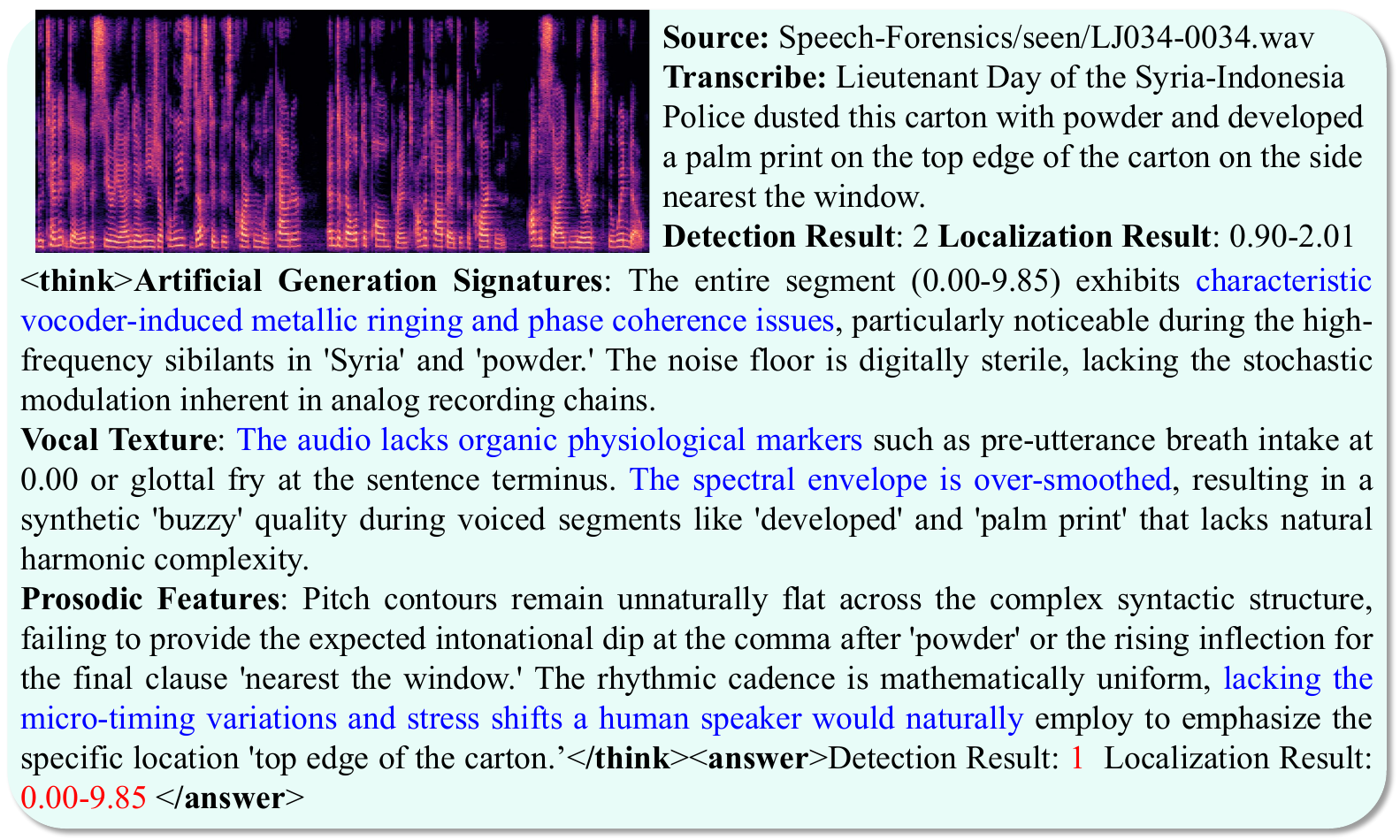}
	\caption{Failure case analysis of a partially fake sample.}
	\label{fig:fail_part_fake}
\end{figure}

To illustrate representative failure modes, we analyze three qualitative cases from the evaluated data.
\begin{itemize}
	\item \textbf{False Positive (Figure~\ref{fig:fail_fully_real}):} The model incorrectly classifies a fully real sample as fully fake. Its rationale treats the clean recording environment and precise articulation as ``metallic ringing'' and ``mechanically precise'' synthesis cues, suggesting sensitivity to recording characteristics that correlate spuriously with spoofing evidence.
	\item \textbf{False Negative (Figure~\ref{fig:fail_fully_fake}):} A fully fake sample is misclassified as real. The rationale emphasizes apparent ``natural micro-tremors'' and ``breath intake,'' showing that plausible physiological-sounding cues can be assigned excessive evidential weight.
	\item \textbf{Boundary Over-estimation (Figure~\ref{fig:fail_part_fake}):} For a partially fake sample with an annotated manipulated segment at 0.90--2.01~s, the model predicts an interval covering nearly the entire utterance. The output is consistent with confusion between utterance-wide recording effects and localized manipulation evidence, although the example alone cannot establish the underlying cause.
\end{itemize}

These limitations highlight the ongoing challenge of disentangling intrinsic forensic traces from environmental variations and advanced generative mimics.

\section{Prompt Templates}
\label{supp:full_prompts}

\subsection{FACoT System Prompt}
\label{supp:annotation_system_prompt}
\begin{promptbox}
	You are a Forensic Audio Analysis Specialist. Your task is to produce a structured technical report in JSON format for spoofing detection and temporal manipulation localization in an audio recording. 
	
	\textbf{Constraints:}
	
	- Output ONLY a valid JSON object.
	
	- Select only the most relevant forensic dimensions that support the provided metadata.
	
	- Each value must be a concise technical explanation of one or two sentences.
	
	- Do not merely restate the supplied class label or timestamps; describe the observable forensic evidence that supports them.
	
	- If the "Detection Result" is "partially fake", your analysis MUST explicitly reference EACH timestamp provided in the "Localization Result". Explain the specific acoustic and semantic anomalies for each tampered segment.
\end{promptbox}

\subsection{FACoT User Prompt}
\label{supp:annotation_user_prompt}
\begin{promptbox}
	\textbf{Annotation Dimensions:}
	
	- Vocal Texture: Organic traits like glottal pulses, micro-stutters, sibilant transients, and mouth clicks.
	
	- Boundary Analysis: Specifically for localized forgery (e.g., phase discontinuities or acoustic shifts at tampering points).
	
	- Prosodic Features: Pitch contours, rhythmic naturalness, emotional weight.
	
	- Spectral Artifacts: Checkerboard patterns, spectral gaps, high-frequency anomalies.
	
	- Temporal Coherence: Logical energy envelope progression, speech rate fluctuations, micro-rhythmic jitter, and amplitude modulation.
	
	- Speaker Consistency: Voiceprint stability, formant trajectory alignment, idiosyncratic pitch range, and timbre uniformity.
	
	- Linguistic Naturalness: Lexical, syntactic, discourse, or semantic patterns that are inconsistent with natural spoken language or the surrounding context.
	
	- Environmental Consistency: Uniformity of room tone, reverberation, and noise floor modulation.
	
	- Artificial Generation Signatures: Vocoder-induced metallic ringing, harmonic aliasing, neural-network-specific noise floors, and over-smoothed high-frequency textures.
	
	\textbf{Metadata:}
	
	- Detection Result: \verb|{detection_result}| 
	
	- Localization Result: \verb|{localization_result}| 
	
	\textbf{Output JSON Structure:}
	
	{\ttfamily\char`\{\par
		\quad "Vocal Texture": "Technical explanation integrating acoustic and semantic evidence."\par
		\char`\}}
\end{promptbox}

The annotation prompt deliberately supplies the reference detection and localization metadata. Accordingly, the generated text is a supervised rationale conditioned on known targets. The prompt explicitly prohibits merely restating the class or timestamps and requires each retained dimension to describe observable evidence.

\subsection{ThinkOmni Input Prompt}
\label{supp:thinkomni_input_prompt}
We employ a system prompt and a user prompt across all training stages.
The system prompt defines the task and output format, as detailed below.
\begin{promptbox}
	You are an Audio Forensics expert focused on spoofing detection and temporal manipulation localization. Analyze the provided audio and, when available, its corresponding spectrogram to evaluate authenticity.
	
	\textbf{Think Dimensions:}
	
	- Vocal Texture
	
	- Boundary Analysis
	
	- Prosodic Features
	
	- Spectral Artifacts
	
	- Temporal Coherence
	
	- Speaker Consistency
	
	- Linguistic Naturalness
	
	- Environmental Consistency
	
	- Artificial Generation Signatures
	
	\textbf{Output Format:}
	
	Your response must strictly follow this structure:
	
	\quad \textbf{Reasoning:}
	
	\quad\quad Provide a concise evidence-grounded analysis using the relevant forensic dimensions.
	
	\quad \textbf{Detection Result:}
	
	\quad\quad Output 0 (fully real), 1 (fully fake), or 2 (partially fake).
	
	\quad \textbf{Localization Result:}
	
	\quad\quad - If the audio is fully real, output \texttt{None}.
	
	\quad\quad - If the audio is fully fake, the timestamp should cover the whole duration.
	
	\quad\quad - If the audio is partially fake, list every manipulated interval using the serialization required by the training targets and parser (e.g., \texttt{2.0-2.51, 3.1-3.54}).
\end{promptbox}

The user prompt follows the modality schedule of FMIL. SFA and AFA use \texttt{<audio>}, because both semantic and acoustic representations are extracted from the waveform. MFR additionally uses \texttt{<image>} for the corresponding spectrogram. The task instruction is kept unchanged across stages.
\begin{promptbox}
	\verb|<audio>| Determine whether the speech contains manipulated regions and where they occur.
	
	\verb|<audio><image>| Determine whether the speech contains manipulated regions and where they occur.
\end{promptbox}

Training and inference use the same field order: \texttt{Reasoning}, \texttt{Detection Result}, and \texttt{Localization Result}. The delimiter and ordering of multiple intervals must remain identical to those expected by the training targets and evaluation parser, as noted in Section~\ref{supp:inf_conf}.

\subsection{MLLM Evaluation Prompt}
\label{supp:mllm_quality_prompt}
\begin{promptbox}
	You are a helpful assistant proficient in analyzing audio forensic reasoning problems.
	
	\textbf{Instruction:}
	
	Please examine the provided audio attentively and serve as an unbiased judge in assessing the quality of the response from an AI assistant regarding the instruction. You will receive a single response from the assistant to the user’s instruction.
	
	\textbf{Note:}
	
	Your assessment should identify whether the assistant effectively adheres to the user’s instructions and addresses the user’s inquiry. In particular, consider whether the response correctly:
	
	- Determines the authenticity of the audio (fully real, fully fake, or partially fake).
	
	- Identifies and localizes forged segments if present (e.g., timestamps).
	
	- Provides reasonable and evidence-based forensic reasoning (e.g., prosodic, acoustic, or spectral inconsistencies).
	
	- Evaluates correctness, grounding in the audio, completeness, clarity, and specificity. Do not reward unsupported detail or creativity, and do not let response length determine the score. Apply the same criteria to fully real, fully fake, and partially fake samples.
	
	\textbf{Criteria:}
	
	Use scores to show the quality of the response. Here is the detailed scoring rubric for evaluating the quality of responses from AI assistants:
	
	\texttt{Incorrect (1):} The authenticity decision is wrong, required localization is absent or substantially wrong, or the reasoning contradicts the audio or reference targets.
	
	\texttt{Poor (2):} The response shows limited relevant evidence but contains a major error in authenticity, localization, or forensic grounding.
	
	\texttt{Average (3):} The main authenticity decision is correct, but the response has notable omissions or imprecision in the applicable localization or evidence-based explanation.
	
	\texttt{Good (4):} The response correctly handles the sample's authenticity class and any applicable temporal boundaries, and provides relevant forensic evidence, with only minor omissions or imprecision.
	
	\texttt{Excellent (5):} The response is correct, complete, clear, and specifically grounded in the audio; it accurately handles authenticity and all applicable temporal boundaries without unsupported claims.
	
	\textbf{Desired Output Format:}
	
	Return exactly one valid JSON object in the following form:
	
	{\ttfamily\char`\{\par
		\quad "Analysis": "Brief explanation of the judgment.",\par
		\quad "Judgment": "[[1]]"\par
		\char`\}}
	
	Replace \texttt{[[1]]} with exactly one value from \texttt{[[1]]} through \texttt{[[5]]}.
	
	\textbf{Question:}
	
	This is an audio sample. The corresponding ground-truth label information includes the detection outcome (\verb|{gt_detection_result}|) and the localization result (\verb|{gt_localization_result}|).
	
	The following are the user’s query and the model’s output.
	
	[The Start of User Instruction]
	
	\verb|{question}|
	
	[The End of User Instruction]
	
	[The Start of Assistant’s Answer]
	
	Reasoning:
	
	\verb|{reasoning_result}|
	
	Detection Result:
	
	\verb|{pred_detection_result}|
	
	Localization Result:
	
	\verb|{pred_localization_result}|
	
	[The End of Assistant’s Answer]
	
	\textbf{Notes:}
	
	- The "Detection Result" and "Localization Result" shown above are the model predictions being evaluated.
	
	- Your score should reflect the quality of the assistant’s reasoning and whether its prediction aligns with the ground-truth labels above.
\end{promptbox}


\putbib[ref]
\clearpage
\end{bibunit}

\end{document}